\documentclass[fleqn,usenatbib]{mnras}

\usepackage{newtxtext,newtxmath}

\usepackage[T1]{fontenc}
\usepackage{ae,aecompl}

\usepackage{graphicx}	
\usepackage{amsmath}	
\usepackage{amssymb}	
\usepackage{subfigure}
\usepackage[normalem]{ulem}

\newcommand{\rev}[1]{{#1}}
\newcommand{\revtwo}[1]{{#1}}
\newcommand{\Fig}[1]{{Fig.\ \ref{fig:#1}}}

\newcommand{\Eq}[1]{Eq.\ \ref{eq:#1}}
\def\rp{R_{\rm P}}
\def\ME{$M_\oplus$}
\def\MEyr{\mathrm{M}_\oplus\,\mbox{yr}^{-1}}
\def\kms{\mathrm{km}\,\mathrm{s}^{-1}}
\def\RH{R_{\rm \mathrm{H}}}
\def\RB{R_{\rm \mathrm{B}}}
\def\ag{{\bf a_{\mathrm G}}}
\def\rr{{\bf r}}
\def\vv{{\bf v}}
\def\ad{{\bf a_{\rm d}}}
\def\vg{{\bf v_{\rm g}}}
\def\dt{\Delta t}

\def\ts{{t_{\rm s}}}

\def\Mps{M_{\mathrm{P*}}}
\def\Mstar{M_\mathrm{star}}
\def\reff{r_\mathrm{eff}}

\newcommand{\OmegaK}{\ensuremath{\Omega_\mathrm{K}}} 

\title[Pebble dynamics and accretion onto rocky planets.]{Pebble dynamics and accretion onto rocky planets.\\ I. Adiabatic and convective models}

\author[{A. Popovas, \AA}. Nordlund, J. P. Ramsey, C. W. Ormel]{
Andrius Popovas,$^{1,3}$\thanks{E-mail: popovas@nbi.ku.dk (AP)}
{\AA}ke Nordlund,$^{1}$\thanks{E-mail: aake@nbi.ku.dk (\AA N)}
Jon P.\ Ramsey,$^{1}$\thanks{E-mail: jramsey@nbi.ku.dk (JPR)}
and Chris W. Ormel,$^{2}$
\\
$^{1}$Center for Star and Planet Formation, the Niels Bohr Institute and the Natural History Museum of Denmark,\\
University of Copenhagen, \O ster Voldgade 5-7, DK-1350 Copenhagen, Denmark\\
$^{2}$Anton Pannekoek Institute, University of Amsterdam, Science Park 904, C4.149, 1098 XH Amsterdam, The Netherlands,\\
$^{3}$Rosseland Centre for Solar Physics, Institute of Theoretical Astrophysics, University of Oslo, P.O. Box 1029 Blindern, N-0315 Oslo, Norway
}

\date{Accepted 2018 June 27. Received 2018 June 14; in original form 2018 January 26}

\pubyear{2018}

\begin{document}
\label{firstpage}
\pagerange{\pageref{firstpage}--\pageref{lastpage}}
\maketitle

\begin{abstract}
We present nested-grid, high-resolution hydrodynamic simulations of gas and particle dynamics in the vicinity of Mars- to Earth-mass planetary embryos. The simulations extend from the surface of the embryos to a few vertical disk scale heights, with \rev{a spatial} dynamic range \rev{of} $\sim\! 1.4\times 10^5$. Our results confirm that ``pebble''-sized particles are readily accreted, with accretion rates continuing to increase up to metre-size ``boulders'' for a 10\% MMSN surface density model. The gas mass flux in and out of the Hill sphere is consistent with the Hill rate, $\Sigma\Omega R_\mathrm{H}^2 = 4\, 10^{-3}\ \MEyr$. While smaller size particles mainly track the gas, a net accretion rate of $\approx 2\,10^{-5}\ \MEyr$ is reached for 0.3--1 cm particles, even though a significant fraction leaves the Hill sphere again. Effectively all pebble-sized particles that cross the Bondi sphere are accreted. The resolution of these simulations is sufficient to resolve accretion-driven convection. Convection driven by a nominal accretion rate of $10^{-6}\ \MEyr$ does not significantly alter the pebble accretion rate. We find that, due to cancellation effects, accretion rates of pebble-sized particles are nearly independent of disk surface density. As a result, we can estimate accurate growth times for specified particle sizes. For 0.3--1 cm size particles, the growth time from a small seed is $\sim$0.15 million years for an Earth mass planet at 1 AU and $\sim$0.1 million years for a Mars mass planet at 1.5 AU.
%
%
\end{abstract}


\section{Introduction}
\label{sec:intro}
We now know that the majority of stars in our galaxy host at least one planet \citep{Cassan2012}. We also know that planets form inside the gaseous and dusty protoplanetary disks (PPDs) that orbit stars during their youth. As it is now possible to spatially resolve the structure of young PPDs (e.g., with the Atacama Large Millimetre/sub-millimetre Array; ALMA), it is becoming apparent that the majority of PPDs have a rich sub-structure of rings, gaps, vortices and spirals (e.g.\ \citealt{vanderMarel2013,isella2013,cassasus2013,broganetal2015_hltau,Andrewsetal2016_twhya,Meru2017}). Recent, spatially resolved observations of young disks with ALMA \citep{broganetal2015_hltau,Andrewsetal2016_twhya}, as well as meteoritic evidence \citep{Bizzarro2017},
furthermore indicate that planet formation starts early. What remains to be answered, however, is how planets grow efficiently enough under realistic conditions to agree with the astronomical observations and meteoritic evidence.

There are currently two scenarios for planet growth which are popular in the literature: First, the planetesimal accretion scenario, in which a planetary embryo is impacted by kilometre-size bodies and their mass is added to the embryo (e.g.\ \citealt{Pollack1996,Hubickyj2005}). Eventually, the embryo becomes massive enough that gravitational focusing \citep{Greenberg1978} becomes important and the embryo enters what is commonly called the `runaway growth' phase. Planetesimal accretion ends when there is no remaining solid material in an embryo's `feeding zone', e.g., when the embryo grows massive enough to gravitationally stir up nearby planetesimals, enhancing their eccentricities and effectively kicking them out of the feeding zone \citep{Ida1993}. The growth of the, now somewhat isolated, protoplanet transitions to the much slower `oligarchic growth' phase \citep{Kokubo1998}. In this phase, the few, largest mass protoplanets grow oligarchically, while the remaining planetesimals mostly remain small. The critical time scale in this context is the lifetime of the PPD, which is of the order of several million years (e.g.\ \citealt{Bell2013}). However, numerical simulations of the planetesimal accretion hypothesis have the problem that they predict that it takes much longer than a PPD lifetime for cores to grow to observed planetary masses (e.g. \citealt{Levison2010,Bitsch2015b}).

The ``pebble accretion'' scenario has, meanwhile, demonstrated promise in being able to accelerate the protoplanet growth process significantly \citep[etc.]{Ormel2010, johansenlacerda2010, Nordlund2011, Lambrechts2012, Morbidelli2012, Lambrechts2014, Bitsch2015b, Chatterjee2014, Visser2016, Ormel2017}. ``Pebbles'', in the astrophysical context, are millimetre to centimetre-sized particles with stopping times, $t_\mathrm{s}$, comparable to their orbital period, $t_\mathrm{s} \sim \OmegaK^{-1}$, where $\OmegaK = \sqrt{GM_*/r^3}$ is the Keplerian frequency, $G$ is the gravitational constant, $M_*$ and $r$ are the mass of and the distance to the central star. Pebbles are likely to form a significant part of the solid mass budget in PPDs, as indicated by both dust continuum observations (e.g.\ \citealt{Testi2003,Lommen2009}) and by the mass fraction of chondrules in chondritic meteorites \citep{Johansen2015,Bollard2017}. \rev{In this scenario, the combined effects of gas drag due to the difference in speed between the slightly sub-Keplerian dust (and gas) and the Keplerian embryo, and gravity ensure that pebbles are captured and settle to the planet}. As the embryo grows, the efficiency of gravitational focusing of pebbles increases, and the effective accretion cross section becomes larger than the embryo itself.

The radius of dominance of the gravitational force of a planet relative to the central star is approximately given by the Hill radius:
\begin{equation}
R_\mathrm{H} = a\sqrt[3]{\frac{M_\mathrm{p}}{3 M_*}},
\label{eq:Hill_radius}
\end{equation}
where $a$ is the semi-major axis of the embryo's orbit, $M_\mathrm{p}$ and $M_*$ are the masses of the embryo and the central star, respectively. Particles of suitable size, passing the embryo even as far away as the Hill radius may be accreted, as has been shown analytically by \cite{Ormel2010}, using test particle integrations on top of hydrodynamical simulations by \cite{Morbidelli2012} and using numerical simulations with particles by \cite{Lambrechts2012}. According to these works, pebble accretion is so efficient that it can reduce planet growth time scales to well within the lifetime of a PPD, even at large orbital distances from the host star.

Herein, we present the first-ever high resolution simulations of gas and pebble dynamics in the vicinity of low-mass, ``rocky'' planetary embryos (for simplicity, often referred to only as `embryos' in what follows) and report on the measured accretion efficiency of pebbles. We consider three embryo masses: 0.95 \ME, 0.5 \ME, and 0.096 \ME, and we specifically choose to study conditions expected for pressure traps associated with inside-out-scenarios of planet formation \citep[e.g.][]{tan_overview_2015}. The embryos are thus assumed to be embedded in disks that have a local pressure maximum at the orbital radius of the embryo.

\rev{The presence of a pressure bump instead of a flat density distribution provides three advantages:
\begin{enumerate}
\item It eliminates the head wind otherwise associated with the slightly sub-Keplerian motion of the gas;
\item It helps to trap particles drifting inwards from larger orbital radii \citep{Whipple1972,paardekooper2006};
\item High resolution observations (e.g. \citealt{broganetal2015_hltau,Andrewsetal2016_twhya}) show disks with rings and gaps, which affect the distribution of solids in the disk. Having the pressure bump allows particles to drift into the pressure maximum, thus mimicking the observed behaviour.
\end{enumerate}
}

The structure of the paper is as follows. In Section \ref{sec:HD_setup} we describe the simulation set up, including initial conditions and their dynamic relaxation to a quasi-stationary state. We then describe, in Section \ref{sec:pebbles}, our treatment of particles, their equations of motion and how they are injected into the simulation domain. In Section \ref{sec:results_gas} we describe the gas dynamics that develops in our simulations and compare our results to other studies. In Section \ref{sec:results_particles} we present our main results regarding particle dynamics and accretion efficiency onto planetary embryos. In Section \ref{sec:convection} we present the first results of simulations of accretion-driven convection in the primordial atmospheres of rocky planets. Finally, in Section \ref{sec:conclusions_discussion}, we summarize our main results, discuss their implications and indicate future work.

\section{Simulation set up}
\label{sec:HD_setup}
This study is carried out using the new DISPATCH framework \citep{Nordlund2017} in a three-dimensional, Cartesian (shearing box) domain, with a set of static, nested patches. We employ a variation of the finite-difference STAGGER solver \citep{nordlund+1994,kritsuketal2011} that, apart from the effects of transformation to a rotating coordinate system detailed below, solves the following set of partial differential equations:
\def\taux{\underline{\underline\tau}}
\def\tauv{\underset{=}{\tau}}
\begin{align}
\frac{\partial {\rm \rho}}{\partial t} & = - \nabla\cdot({\rm \rho}\mathbf{u}) ;\label{eq:continuity} \\
  \frac{\partial {\rm \rho}\mathbf{u}}{\partial t}  & = -\nabla\cdot({\rm \rho}\mathbf{u}\mathbf{u} + \tauv) -\nabla (p+p_a) - \rho\nabla\Phi;\label{eq:momentum} \\
  \frac{\partial s}{\partial t}  & = -\nabla\cdot(s\mathbf{u}),
\label{eq:entropy}
\end{align}
on a staggered mesh, where $\rm \rho$ is the gas density, $\mathbf{u}$ is the gas velocity, $p$ is the gas pressure, $p_a$ is a stabilising artificial pressure term, $\tauv$ is a viscous stress tensor, $s = \rho \log(p/{\rm \rho}^\gamma) / (\gamma - 1)$ is the entropy density per unit volume, $\gamma$ is the ratio of specific heats and $\Phi$ is the gravitational potential.  The viscous stress tensor is taken to be
\begin{equation}
\tauv = \tau_{i,j}= \nu_1\Delta x (c_s+u)\rho\left(\frac{\partial u_i}{\partial x_j}+\frac{\partial u_j}{\partial x_i}\right),
\label{eq:stress_tensor}
\end{equation}
where $\Delta x$ is the grid resolution, $c_\mathrm{s}$ is the sound speed, and the coefficient $\nu_1$ is a small fraction of unity.  The artificial pressure is computed as:
\begin{equation}
p_a=  \nu_2\rho \min(0,\Delta x (\nabla\cdot\mathbf{u}))^2,
\label{eq:artificial_pres}
\end{equation}
where $\nu_2$ is a coefficient $\sim 1$ \rev{\citep{kritsuketal2011}}.

This allows a discretisation that explicitly conserves mass, momentum and entropy.  The viscous stress tensor stabilizes the code in general, while the artificial pressure term ensures that shocks are marginally resolved.  

The equation of state (EOS) considered herein is that of an ideal gas with adiabatic index $\gamma = 1.4$ and molecular weight $\mu = 2$. Although not considered here, in a forthcoming paper we include radiative energy transfer and adopt a more realistic, table-based EOS (Popovas et al., in prep.).

We use code units where length is measured in Earth radii, mass is measured in Earth masses, and velocity is measured in units of 1 km s$^{-1}$. \rev{Hence our time unit is $R_\oplus/1 \kms$ = 6371 seconds, corresponding to roughly 5,000 time units per year.}

Partly because the time step is determined locally in each patch, and partly because of the small number of cells per patch ($33^3$ in this case), the hydro solver operates near its optimal speed---updating on the order of 1.4 million cells per core per second---we performed these simulations using only a single, 20-core Intel Ivy Bridge node for each experiment.

\subsection{Grid set up}
\label{subsec:grid}
We use nested sets of Cartesian patches centred on the embryo. With the embryo as the origin of our coordinate system, the $x$-axis is parallel to the cylindrical radial direction and points away from the central star, while the $y$- and $z$- axes are oriented along the azimuthal and vertical directions, respectively.  The $+y$-axis lies along the direction of orbital rotation. The embryo is assumed to move in a circular orbit. Patches are arranged in a nested, hierarchical `Rubik's Cube' arrangement,with 216 patches arranged into a $6\times 6\times 6$  cube. The central 8 patches are then split recursively into $6\times 6\times 6$ child patches, which have a three times finer resolution than the previous level. This is repeated for up to 7 levels of refinement (8 levels for the smallest embryo mass), while the coarsest set is duplicated 5 times in the $y$-direction. The size of the innermost, central, patch, exceeds the embryo radius $\rp$ by 10-30\%. With 7 levels in the hierarchy, the outer scale in the vertical and radial directions is then $\sim 5000 \rp$, which is about 4 times a typical disk scale height for an Earth-mass planet, and $\sim$20x larger than the Hill radius. In the $y$-direction, the model extends to $\sim\! 25000 \rp$. Each patch contains $33\times 33\times 33$ cells, leading to a maximum resolution in our simulations, near the embryo's surface, of $\approx 3\% \rp$. \rev{This resolution is  higher than has been used in previous works \citep[e.g.][]{dangelo2013,Cimerman2017,Lambrechts2017}. Since the numerical viscosity is proportional to the cell size (cf.\ \Eq{stress_tensor}), the numerical viscosity in our simulations is also correspondingly smaller than in these previous studies\footnote{\rev{As demonstrated in \cite{kritsuketal2011}, the effects of numerical viscosity are generally not larger in codes that use explicit numerical viscosity than in codes that rely on viscous effects implicit to the methods.}}.}

\subsection{Initial conditions}
\label{subsec:initial_conditions}
We consider a local shearing flow in the reference frame that co-moves with the embryo at the local Keplerian frequency, $\OmegaK$. We ignore the slight curvature of the partial orbit but include the Coriolis force resulting from the orbital motion. The origin of the co-moving frame is, depending on the simulation run, placed at either 1 or 1.5 astronomical units (AU). The initial gas velocity is then given by:
\begin{equation}
\mathbf{u}(t = 0) =  - \left(\frac{3}{2} + \zeta_p\right)\OmegaK x \mathbf{\hat{y}},
\label{eq:accelerations}
\end{equation}
where $\zeta_p$ is a pressure trap parameter, defined below. In this study, we do not consider a systematic accretion flow towards the central star, nor the headwind that an embryo experiences when the gas is slightly sub-Keplerian. In analytically constructed accretion disks, with pressure decreasing systematically with radius, the headwind is typically on the order of one percent of the Keplerian velocity, but when an embryo is embedded in a local pressure bump there is no headwind. A systematic accretion flow would allow the smallest size particles to escape the pressure trap, and would thus tend to increase the fraction of large size particles, but since we are mainly concerned with measuring the accretion {\it efficiency} of different size particles, the possible evolution of the grain size distribution is not of direct importance for our study.

The background vertical density stratification is set by the balance between the stellar gravity and the gradient of the gas pressure:
\begin{equation}
\rho(z) = \rho_0 \, \exp\left(-{\frac{\OmegaK^2 z^2 + 2\zeta_p x^2}{2p_0/\rho_0}}\right),
\label{eq:vertical_stratification}
\end{equation}
where $\rho_0$ and $p_0$ are the disk midplane density and pressure, respectively, and are estimated from the 2D radiative disk models of \citet{Bitsch2015}. The $\zeta_p$ term gives rise to a pressure gradient that balances the extra shear introduced in \Eq{accelerations} above. \rev{A pressure bump produces a more self-contained region in the vicinity of the embryo's orbit, which retains the density wake produced by the embryo, preventing it from causing a forceful numerical bounce at the boundaries. This is also mitigated by the wave-killing boundaries that we use (see below), but a less forceful wake means a smaller region at the boundaries needs to be artificially damped.}

With typical dust opacities\footnote{Chondritic meteorites typically have a fair fraction of ``matrix'', which corresponds to essentially unprocessed dust. This demonstrates that, at least during the epoch when chondritic parent bodies formed, dust provided substantial opacity per unit mass in the proto-Solar disk.} on the order of 0.1 - 1 cm$^2$ g$^{-1}$, disks with surface densities similar to the Minimum Mass Solar Nebula \citep[MMSN;][]{Hayashi81} are optically thick in the vertical direction, since the MMSN nominally has 1700 g/cm$^2$ at 1 AU. If early hydrostatic atmospheres that surround planetary embryos are assumed to be nearly adiabatic  one can predict their approximate vertical structure from knowledge of only the density, $\rho_0$, and temperature, $T_0$, at the outer boundary. The pressure at the outer boundary of the atmosphere is assumed to match the local pressure of the surrounding disk. Therefore, for reasonable initial conditions, we approximate an adiabatic atmosphere profile outside-in to the embryo. From the EOS we  have the pressure at the outer boundary:
\begin{equation}
P_0 = \frac{\rho_0 k_\mathrm{B} T_0}{\mu m_\mathrm{u}},
\label{eq:pressure_outer_boundary}
\end{equation}
where $k_\mathrm{B}$ is the Boltzmann constant and $m_\mathrm{u}$ is the atomic mass unit.  For each layer of the atmosphere, $r_i$, we iterate over the set of equations:
\begin{align}
\rho_i & = \frac{P_i\mu m_\mathrm{u}}{k_\mathrm{B} T_i};\label{eq:hse_rho}\\
\Delta \ln P_i & = \Delta \ln r_i \ \frac{G  M_\mathrm{p}}{2} \left(\frac{\rho_{i+1}}{P_{i+1}r_{i+1}}+\frac{\rho_{i}}{P_{i}r_{i}}\right);\label{eq:hse_pressure}\\
{\Delta \ln T_i} & = {\Delta \ln P_i}\frac{\gamma-1}{\gamma},\label{eq:hse_temperature}
\end{align}
where $\Delta \ln r_i=2(r_{i+1}-r_i)/(r_{i+1}+r_i)$. The temperature and pressure for each layer are then simply $T_i = T_{i+1}e^{\Delta \ln T_i}$ and $P_i = P_{i+1}e^{\Delta \ln P_i}$, respectively. 
We use 500 logarithmically-spaced layers between half the embryo radius and the Hill radius. We then do a polynomial interpolation between these layers to map the density, pressure and temperature to each cell in the simulation box in order to construct the initial conditions in the vicinity of the embryo. 

\begin{figure}
\centering
    \includegraphics[width=0.85\columnwidth]{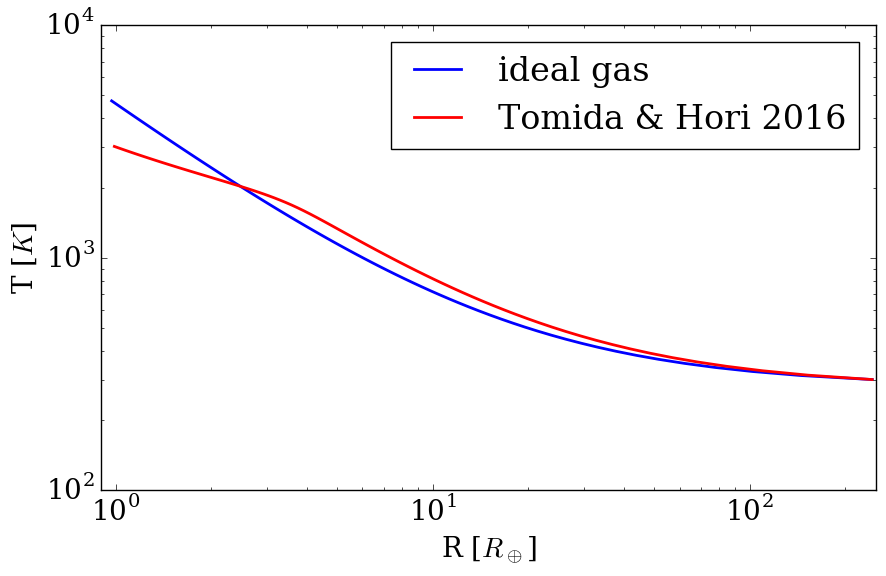}
    \caption{Radial temperature profile for a planetary embryo atmosphere using an ideal gas EOS ($\gamma=1.4$, blue curve) and a realistic, tabular EOS (red curve).}
    \label{fig:eos}
\end{figure}

For the purposes of this study, an ideal gas EOS ($\gamma=1.4$) is sufficient. As demonstrated by \Fig{eos}, the thermal profiles of the atmosphere surrounding an $M=0.95 M_\oplus$ embryo given by an ideal gas EOS and a realistic, tabular EOS (\citealt{Tomida2013}, updated by \citealt{Tomida2016}) are only significantly different close to the surface of the embryo. Gas flows in the vicinity of the Hill sphere, as well as embryo particle accretion rates, are insensitive to these differences.

With constant opacities and an internal heating source (due to the accretion of solids) such an atmosphere is convective, and hence has a nearly adiabatic structure \citep{stevenson1982_giantplanets}. The atmosphere might have a radiative outer shell --- cf.\ the works by \citet{pisoetal2015_mincoremass} on giant planets and by \citet{inaba2003} for a terrestrial mass planet --- but this shell would not substantially change the physical structure of the atmosphere, and ignoring it is a reasonable approximation when constructing the initial conditions. In forthcoming work, however, we include radiative energy transfer in our simulations, and there the initial conditions take this radiative shell into account. 

The effective, dynamical boundary between the disk and the embryo atmosphere occurs at distances from the planetary embryo somewhere in the range between the Hill radius and the Bondi radius,
\begin{equation}
R_\mathrm{B} = \frac{G M_\mathrm{p}}{c_\mathrm{iso}^2},
\label{eq:Bondi_radius}
\end{equation}
where $c_\mathrm{iso}^2$ is the isothermal sound speed. In fluid dynamics, the Bondi radius has no particular dynamical significance, other than being the radius where the pressure scale height of the atmosphere becomes comparable to the distance from the embryo, and the radius where trans-sonic bulk gas speeds would become unbound.

Note that the mass in such an adiabatic atmosphere is completely determined (apart from the equation of state) by the pressure and temperature at the outer boundary, and that if the external pressure is reduced the resulting mass is smaller.  This implies that mass must be lost from such atmospheres when the disk evolves to lower densities \citep{Nordlund2011,Schlichting2015,Ginzburg2016,Rubanenko2017}.

\subsubsection{Parameter space}
\label{subsub:parameters}
In this study, we consider 8 cases: 3 different embryo masses ($M_\mathrm{p}$ = \{0.95, 0.5, 0.096\} \ME) at 2 distances from the central star ($a$ = \{1.0, 1.5\} AU), with different disk midplane densities ($\rho_0$ = \{$1\ 10^{-10}$, $4\ 10^{-11}$\} g cm$^{-3}$) and temperatures ($T_0$ = \{230, 300\} K), as well as with/without a pressure bump ($\zeta_p$ = \{0.0, 0.05, 0.1\}). The basic parameters for all cases are summarised in Table \ref{tab:initial_conditions}.

The radii of the embryos are estimated using the mass-radius formula from \citet{Zeng2016}:
\begin{equation}
\left(\frac{\rp}{R_\oplus}\right) = (1.07 - 0.21 \ CMF)\left(\frac{M}{M_\oplus}\right)^{1/3.7},
\label{eq:massradius}
\end{equation}
where $CMF$ stands for core mass fraction. We take $CMF = 0.325$, although the exact value is not important; taking $CMF=0.0$ only changes the radius of the embryo on the order of 5\%. 

\begin{table*}
\centering
{%
\caption{\label{tab:initial_conditions} Basic parameters for the different simulation runs. See the text for more details.}
\begin{tabular}{lcccccccccc}
Run     & $M_\mathrm{p}$ [$M_\oplus$]  & $R_\mathrm{p}$ [$R_\oplus$]  & $a$ [AU]   & $T_0$ [K]  & $\rho_0$ [g cm$^{-3}$] & $\frac{\OmegaK}{2\pi}$  [yr$^{-1}$] & $\zeta_p$ & $x_\mathrm{damp}$  & $\Delta x_\mathrm{min}$ [$R_\oplus$] & box [$R_\oplus$] \\
\hline
\hline
\texttt{m095t00}  & 0.95  & 0.988 & 1.0 & 300 & $1 \ 10^{-10}$ & 1.0 & 0.0 & 700 & 0.037 & 5186 x 5186 x 25930 \\
\texttt{m095t05}  & 0.95  & 0.988 & 1.0 & 300 & $1 \ 10^{-10}$ & 1.0 & 0.05 & 700 & 0.037 & 5186 x 5186 x 25930 \\
\texttt{m095t10}  & 0.95  & 0.988 & 1.0 & 300 & $1 \ 10^{-10}$ & 1.0 & 0.1 & 700 & 0.037 & 5186 x 5186 x 25930 \\
\hline
\texttt{m05t00}   & 0.5  & 0.83 & 1.0 & 300 & $1 \ 10^{-10}$ & 1.0 & 0.0 & 500 & 0.031 & 4360 x 4360 x 21801 \\
\texttt{m05t05}   & 0.5  & 0.83 & 1.0 & 300 & $1 \ 10^{-10}$ & 1.0 & 0.05 & 500 & 0.031 & 4360 x 4360 x 21801 \\
\texttt{m05t10}   & 0.5  & 0.83 & 1.0 & 300 & $1 \ 10^{-10}$ & 1.0 & 0.1 & 500 & 0.031 & 4360 x 4360 x 21801 \\
\hline
\texttt{m01t00}   & 0.096  & 0.532 & 1.5 & 230 & $4 \ 10^{-11}$ & 0.5433 & 0.0 & 400 & 0.023 & 9773 x 9773 x 48866 \\
\texttt{m01t10}   & 0.096  & 0.532 & 1.5 & 230 & $4 \ 10^{-11}$ & 0.5433 & 0.1 & 400 & 0.023 & 9773 x 9773 x 48866 \\
\hline
\texttt{m095t10-conv}  & 0.95  & 0.988 & 1.0 & 300 & $1 \ 10^{-10}$ & 1.0 & 0.1 & 700 & 0.037 & 5186 x 5186 x 25930 \\
\\
\end{tabular} }
\end{table*}

\subsection{External forces (accelerations)}
\label{subsec:force_mod}
External forces, arising from the transformation to a rotating coordinate system, and from linearising the force of gravity due to the central star, are added to the hydrodynamics as source terms:
\begin{equation}
\frac{\partial \mathbf{u}}{\partial t} = -\frac{GM_\mathrm{p} (x \mathbf{\hat{x}} + y \mathbf{\hat{y}} + z \mathbf{\hat{z}})}{r^3} \ + 3 \OmegaK^2 x \mathbf{\hat{x}} \ - \OmegaK^2 z \mathbf{\hat{z}} \ - 2 \OmegaK \mathbf{\hat{z}} \times \mathbf{u},
\label{eq:external_force}
\end{equation}
where $r$ is the distance from the embryo. The first term in \Eq{external_force} is the two-body force from the embryo. The second term is the tidal force and has two contributions: the centrifugal force and the differential change of the stellar gravity with radius. Vertically, there is a contribution from the projection of the force of gravity onto the z-axis (the third term). The last term is the classical Coriolis force.

To avoid the singularity of the two-body force, rather than using a `softening' of the potential (e.g. \citealt{Ormel2015,Lambrechts2017,Xu2017}), we truncate it via $r = \max(r, 0.1 \rp)$. Truncation is better than softening, since the gas outside the radius of the embryo feels the correct force, and the spherical boundary conditions (described below) negates the effect of truncation for the innermost cells, well within the radius of the spherical boundary condition/embryo.

\subsection{Boundary conditions}
\label{subsec:boundary_conditions}
Using test simulations, we have ensured that our boundary conditions remain stable over long periods of integration time and under rather extreme conditions (e.g.\ highly supersonic flows, low densities). The ``azimuthal'' $y$-boundaries are set to be periodic, whereas the vertical $z$- and ``radial'' $x$-boundaries are set as follows:
\begin{itemize}
\item Zero-gradient conditions for the entropy per unit mass.
\item The $x$- and $z$-velocity components are set symmetric with respect to the boundaries, while the $y$-component is extrapolated in the $x$-direction and $z$ directions.
\item The density boundary condition\rev{s} in the $x$-direction is constructed from assuming hydrostatic balance:
\begin{equation}
\rho_x = \rho_{x_i} \ \exp\left(-\frac{\OmegaK^2 \zeta_p}{T_0}(x^2-x_i^2)\right),
\label{eq:boundary_rho}
\end{equation}
where $\rho_x$ is the density in the boundary at location $x$ and $\rho_{x_i}$ is the density at location $x_i$, which corresponds to the last cell inside a patch's active domain.
\item If a pressure trap is not present we assume a vanishing derivative of the density in the $x$-direction.
\item In the vertical direction, we apply a similar condition:
\begin{equation}
\rho_z = \rho_{z_0} \ \exp\left(-\frac{\OmegaK^2}{2T_0}(z^2-z_0^2)\right),
\end{equation}
where $\rho_z$ is the density in the boundary at location $z$ and $\rho_{z_0}$ is the density at location $z_0$, which again corresponds to the last cell inside the active domain.
\end{itemize}
We additionally apply the ``wave-killing'' prescription of \citet{devalborro2006} to reduce reflections at the $x$-boundaries. It is only applied over a specific range, given by $x_\mathrm{damp}$ (column 9 in Table \ref{tab:initial_conditions}).

\subsubsection{Spherical boundaries}
\label{subsubsec:spherical_bnd}
\begin{figure}
  \begin{minipage}[c]{0.48\columnwidth}
    \caption{Treatment of spherical boundary conditions on a staggered mesh. Blue and red squares represent horizontal and vertical mass fluxes outside the boundary, which evolve freely, whereas turquoise and magenta squares represent mass fluxes inside the boundary, which are set to zero. Green squares represent density points that are subject to change, while yellow squares show density points that remain constant throughout the simulation.}\label{fig:spherical_bnd}
  \end{minipage}\hfill
\begin{minipage}[c]{0.5\columnwidth}
    \centering
    \includegraphics[width=0.99\columnwidth]{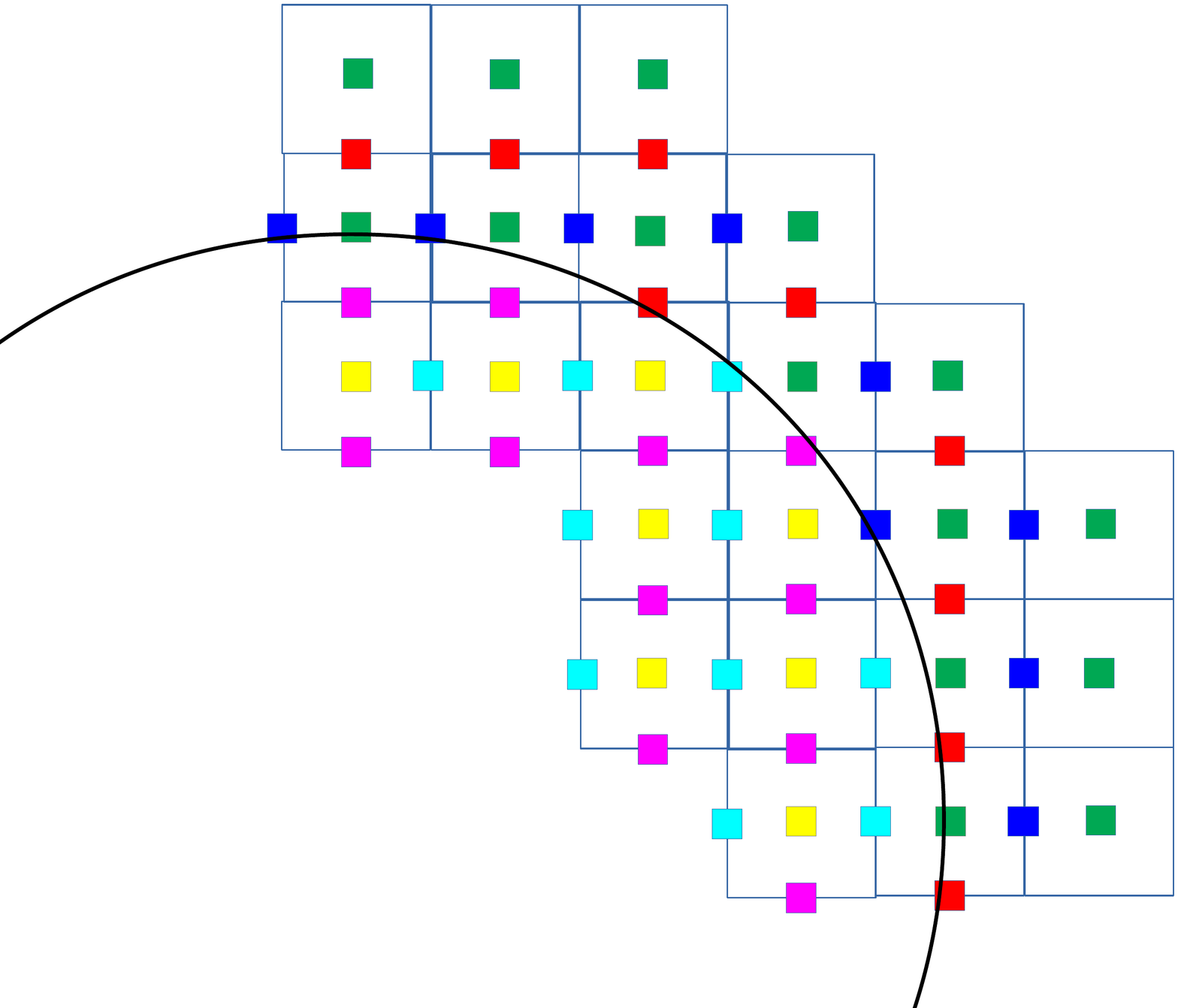}
  \end{minipage}
\end{figure}

In these simulations we use ``jagged'', no-slip boundaries to approximate the embryo surface. \Fig{spherical_bnd} shows a 2D representation of these boundary conditions on a staggered mesh. Density and entropy are allowed to evolve freely everywhere; only the mass flux is constrained. We deem a set of cell faces that approximate the spherical surface to belong to the boundary. If the cell face is inside the spherical boundary, the mass flux is set to zero, otherwise it is allowed to evolve freely. Hence, density cannot change inside the boundary and cells that have faces belonging to the boundary react exactly as they should: inflows towards the boundary produce a positive $\partial \rho /\partial t$ at the boundary. The corresponding increase in pressure causes an increase in the outward directed pressure gradient at the last `live' mass flux point, as expected at a solid boundary. This permits the system to find an equilibrium and react to perturbations from it, and since there are never any inflows and outflows through the boundary, mass is conserved.

\subsection{Relaxation of initial conditions}
\label{subsec:relaxation}
We allow the system to relax for 30 000 time units \rev{(which corresponds to about 6 years),} in order to ensure that dynamically consistent conditions are attained. Indeed, as discussed below, tests demonstrate that it takes on the order of only one orbit for the initial conditions to relax.

The density stratification well inside the Hill sphere is almost hydrostatic, but with deviations and fluctuations on short time scales, and it is thus not necessary to include the detailed dynamics of this region while larger structures are relaxing. Therefore, we begin our experiments by fixing the hydrostatic structure there, and set the spherical boundary to a radius of 27 embryo radii. We then `release' layers (add an extra level of refinement, reduce the radius of the spherical boundary, turn on the dynamics outside of the new spherical boundary) at successively smaller radii (9, 3 and finally, 1 embryo radii). The `releases' are typically done at times = 10 000, 20 000 and 26 000 in code units. Successive releases are increasingly more expensive due to the smaller cell size and larger sound speeds, but need less time in code units to adjust to the surrounding environment (for essentially the same reasons, i.e., smaller scales and larger sound speeds). This procedure saves significant amounts of computing time during the initial relaxation phase.

\begin{figure*}
  \centering
  \subfigure[Initial conditions at time = 0.]{\includegraphics[width=0.95\columnwidth]{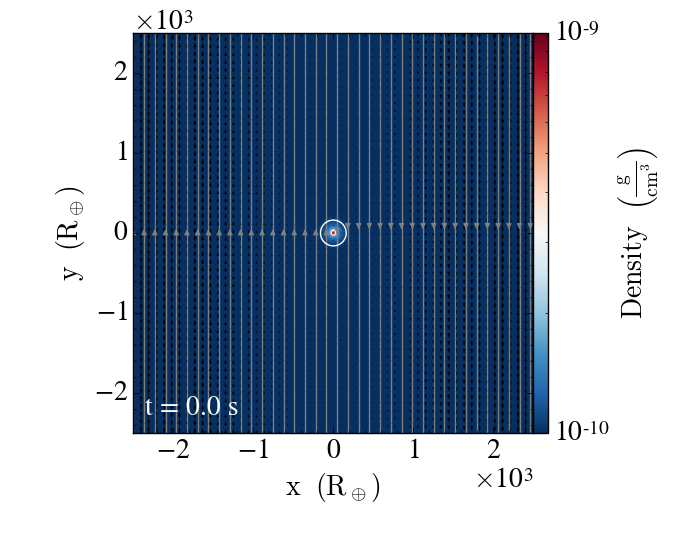}}
  \subfigure[Relaxed simulation at time = 30 000]{\includegraphics[width=0.95\columnwidth]{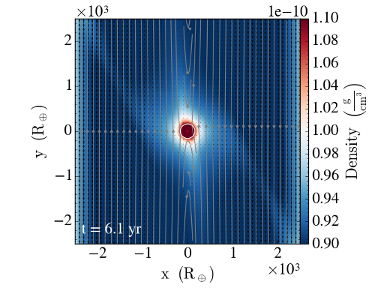}}
  \caption{Relaxation of initial conditions for the \texttt{m095t00} run. The panels show density slices and velocity streamlines at the midplane, with almost the entire extent of the simulation box in the $x$-direction and 1/5 of the way in the $y$-direction. Horseshoe orbits are clearly visible in panel (b). Panel (b) is plotted with a highly enhanced contrast (the colour scale spans only 20\% in density) to show the wakes and density increase at the orbital radius in the $y$-direction. The white and red circles (barely visible in these figures) denote the Hill radius and the isothermal Bondi radius of the embryo, respectively. The embryo itself is smaller than a pixel.}
  \label{fig:ic_relax}
\end{figure*}

\begin{figure*}
  \centering
  \subfigure[Initial conditions at time = 0.]{\includegraphics[width=0.95\columnwidth]{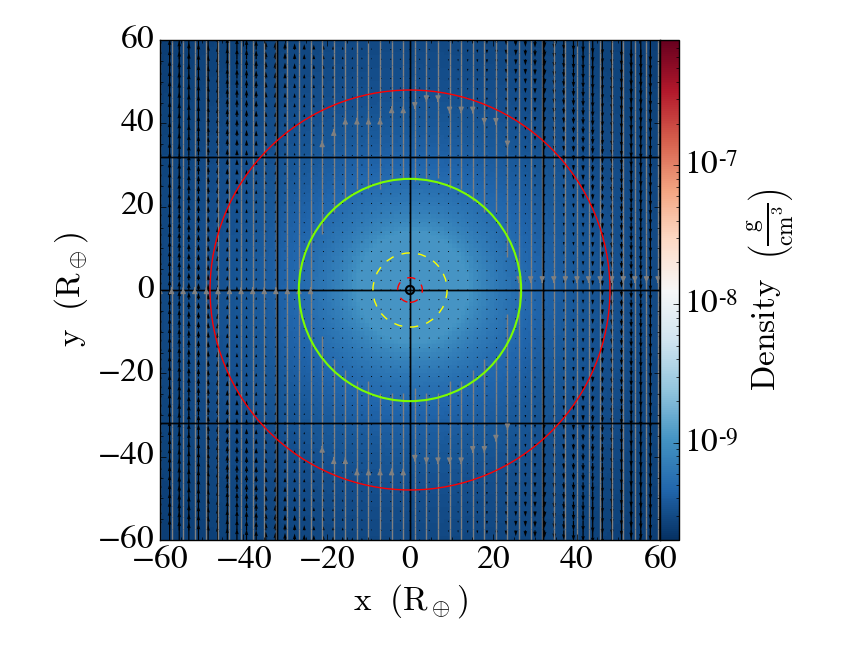}}
  \subfigure[Relaxed simulation at time = 30 000]{\includegraphics[width=0.95\columnwidth]{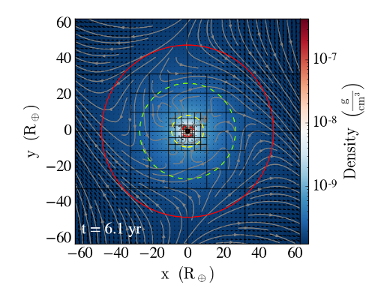}}
  \caption{Relaxation of initial conditions for the \texttt{m095t00} run. Intermediate scale density slices at the midplane. The large red circle is the Bondi radius. The solid green circle in panel (a) shows the current `active' radius, while subsequent dashed circles show where future `releases' will place a temporary embryo surface. The tiny black circle shows the actual radius of the embryo, which is reached after the final release.}
  \label{fig:ic_relax_zoom}
\end{figure*}

After the last release, the simulations are allowed to relax for an additional 4000 time units before they are deemed ready for the next phase. Figures \ref{fig:ic_relax} and \ref{fig:ic_relax_zoom} show large-scale and zoomed-in density slices at the midplane for run \texttt{m095t00} at $t$ = 0 and after a relaxed state is reached at $t$ = 30 000. Panels on the right show the relaxed state with well established horseshoe orbits, density wakes, and slight density perturbations ahead/behind of the embryo at the same orbital distance (discussed in more detailed below). 

\Fig{ic_relax_zoom} is a zoom-in of the same run with the Bondi radius more clearly visible (largest red circle); the Rubik's cube nesting of grids is also visible. The dashed lines indicate the successive releases of the radius of the embryo (green - 27, yellow - 9, light red - 3 embryo radii). The tiny black circle corresponds to one embryo radii. The right panel shows the simulation when fully relaxed and, via streamlines, demonstrates that the disk gas penetrates deeply into $R_\mathrm{B}$, down to $\approx\! 20R_\mathrm{P}$. At even smaller radii the atmosphere is nearly static, with motions that have a weak rotational tendency (see below).

\begin{figure}
\centering
    \includegraphics[width=\columnwidth]{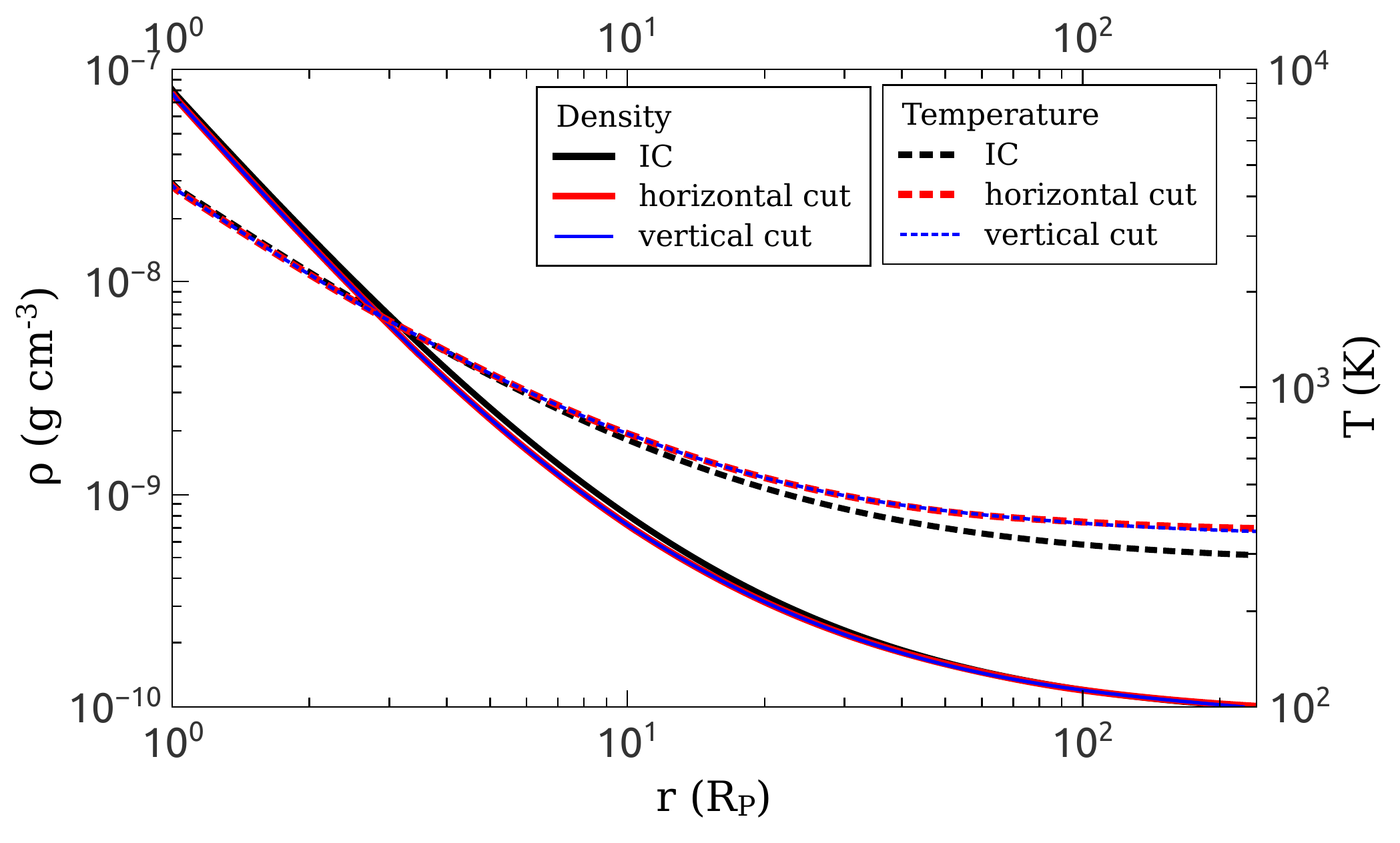}
    \caption{Physical structure of the primordial atmosphere for run \texttt{m095t00} at $t = 0$ (initial conditions; IC) and $t$ = 30 000 = 6.1 years (fully relaxed). The horizontal and vertical cuts overlay each other nearly perfectly.}
    \label{fig:hse}
\end{figure}

\Fig{hse} shows the physical structure of the primordial atmosphere at both time = 0 and 30 000. The cuts are done along the $x$-axis in the midplane and along the $z$-axis at $x = y = 0$. As can be seen, the initial state is very close to the final state of the atmosphere, and there are only negligible variations for different cuts. Although the density and temperature profiles have changed slightly, the changes are minimal and the use of these initial conditions thus saves a lot of relaxation time (compared to, e.g., a sudden insertion of the planet's gravitational potential, which would require waiting tens of orbits with the maximum level of refinement for the dynamics to relax). Although the indnermost parts of the atmosphere still retain some weak random motions, by time = 30 000, the run has nearly reached a steady-state, when injection of particles can commence. Note that, as has been pointed out by \cite{Ormel2015b}, a completely steady, static state will likely never be reached in the vicinity of the embryo.

\section{Particles}
\label{sec:pebbles}
\rev{After relaxation of the fluid motions and density distribution for 30,000 time units (about 6 years) we add particles for the purpose of tracing the motions of selected sub-samples and reveal, for example, the typical paths of different physical size particles and to be able to measure the rate of accretion onto the planetary embryos.  In these simulations, we consider macro-particles (sometimes called super-particles; \citealt{Rein+2010,Ebisuzaki2017}), each of which represents a swarm of real particles, with a given size and mass.
We follow the particle motions, taking into account external forces and gas drag, until they are accreted onto the planet embryo, or exit the domain through the $x$-boundaries.  We consider particles as `accreted' when their radial distance to the centre of the planet is smaller than the radius of the embryo.}

\rev{In order to have statistical coverage everywhere,} the initial spatial distribution of macro-particles is chosen to be proportional to the local gas density. The actual dust-to-gas ratio is likely characterized by a strong settling towards the midplane, while the dust-to-gas surface density ratio should be essentially constant in the upstream flow entering the Hill sphere. However, rather than having to make assumptions about the settling, we instead analyse sub-populations of our initial distribution, tagging and following for example only the particles that initially reside within a given distance from the midplane.

We use macro-particles with corresponding real particle sizes normally ranging from 10 $\rm \mu$m to 1 cm, with a constant number of macro-particles per logarithmic size bin. However, note that we never integrate over size. The constant number per bin was chosen to provide good statistical coverage for all sizes when investigating the transport of particles in narrow size bins.

We include 1 cm particles for completeness, even though such large particles are typically found only in CB chondrites \citep{Friedrich2015}, partly because there is no guarantee that the size distribution of chondrules in chondrites spans the entire size range that was available at the time of pebble accretion onto planet embryos. We also made complementary experiments with even larger size particles (up to 1 m), to explore saturation of accretion rates with size.

The shape of the relative size distribution will be modified near the embryo, because large particles accrete more rapidly onto the embryo, while being replenished at the same rate as smaller size particles. The resulting change of the relative size distribution is a result of this study, and does not need to be anticipated in the initial conditions.

Note that, since we do not consider particle growth and destruction in this study, nor the back reaction of the particles on the gas, we are free to renormalise the the actual size distribution of particles {\em a posteriori} by changing the interpretation of the statistical weight factors of the macro-particles. We can thus match any given initial actual size distribution, and derive what the final distribution would be. We also note that since the span in space and time of our models is very limited (relative to disk extents and lifetimes), ignoring coagulation and shattering is an excellent approximation; such effects may instead be taken into account by modifying -- {\em a posteriori} as per the remarks above -- the initial particle distribution.

\subsection{Equations of motion}
\label{subsec:eqs_of_motion}
The acceleration of macro-particles,
\begin{equation}
\frac{\rm{d}\mathbf{u}_p}{\rm{d}t} = \ag + \ad,
\end{equation}
is governed by $\ag$, the differential forces originating from the planet, the star and the Coriolis force, which, in analogy with Eq. \ref{eq:external_force}, are
\begin{equation}
\ag = -\frac{GM_\mathrm{p} (x \mathbf{\hat{x}} + y \mathbf{\hat{y}} + z \mathbf{\hat{z}})}{r^3} \ + 3 \OmegaK^2 x \mathbf{\hat{x}} \ - \OmegaK^2 z \mathbf{\hat{z}} \ - 2 \OmegaK\mathbf{\hat{z}} \times \mathbf{u},
\label{eq:pebble_grav}
\end{equation}
and the acceleration from the gas drag, 
\begin{equation}
\ad = \frac{\mathbf{u}_\mathrm{gas}-\mathbf{u}_\mathrm{part}}{t_\mathrm{s}},
\end{equation}
which is a product of the relative velocity of the particle, $\mathbf{u}_\mathrm{part}$, with respect to that of the gas, $\mathbf{u}_\mathrm{gas}$, and the inverse of the stopping time, $t_s$, which, in the Epstein regime is:
\begin{equation}
t_\mathrm{s} = \frac{\rho_\bullet s}{\rho v_\mathrm{th}},
\label{eq:tstop_Epstein}
\end {equation}
where $\rho_\bullet$ is the solid density (taken to be 3 g cm$^{-3}$) and $v_\mathrm{th}$ is the mean thermal velocity of the gas. We further assume that particles are spherical.

For simplicity and speed, we parametrise the stopping time as:
\begin{equation}
\left(\frac{\ts}{1\, \mathrm{yr}}\right) = 
\left(\frac{s}{1 \mathrm{cm}}\right) 
\left(\frac{c_\mathrm{d}}{\rho}\right),
\label{eq:tstop_inverse}
\end {equation}
where $c_\mathrm{d} = 10^{-12}$ g cm$^{-3}$ is a normalisation constant, defined such that in the sub-sonic Epstein regime, the stopping time is 1 year for a 1 cm particle in a gas density of $10^{-12}$ g cm$^{-3}$. This estimates the stopping time in the disk and outer parts of the Hill sphere sufficiently well, while ignoring the additional decrease of the stopping time due to the higher temperature near the surface of the embryo. Particles that reach such depths of the potential well are in any case doomed to accrete, so nothing is lost by ignoring the temperature dependence, and doing so reduces the cost to compute the stopping time.

Although this approximation may overestimate the stopping time slightly, the precise value is not critical, since there is in any case an uncertain factor that depends on the shape and porosity of the particles, and a different value can be compensated for by a corresponding shift of the assumed particle distribution.

In a similar vein, there is no need to expand the parameter space with different assumed disk densities, since that too corresponds to a rescaling of the particle stopping time, and a corresponding rescaling of the disk and adiabatic atmospheric structures with a constant factor in density.

Because of its dependence on gas density and particle size, the stopping time can vary over many orders of magnitude. Given the time step, $\dt$, that is necessary based on the velocity and acceleration of gravity, $\ag$, one can thus have both $\dt \ll \ts$ and $\dt \gg \ts$. The latter case means that the differential equation governing the particle path becomes {\em stiff}. 

To handle this, particle positions and velocities are updated with a modified kick-drift-kick leapfrog scheme (e.g.\ \citealt{Dehnen2011}), reformulated to give the correct relative speed also in the asymptotic limit when the time step is much larger than the stopping time. Instead of sub-cycling, which is computationally expensive, we use the method that is commonly referred to as ``forward time differencing'', which ensures that the solution has the correct asymptotic behaviour, both when the stopping time is very long and very short, and that the behaviour is also nearly the correct one in intermediate cases.

With forward time differencing, the velocity update, assuming for now only an acceleration due to the drag force, $\ad$, may be written:
\begin{equation}
 \vv(t+\dt) = \vv(t) + \Delta\vv = \vv(t) + \dt \,\ad (t+\dt) ~,
 \label{eq:pebble_v_upd}
\end{equation}
where the acceleration is time-centred at $t+\dt$. The resulting update relation is:
\begin{equation}
 \Delta\vv = \dt \,\ad (t+\dt) = - (\dt/\ts) \, (\vv+\Delta\vv-\vg(t+\dt))  ~,
 \label{eq:pebble_deltav_upd}
\end{equation}
with solution:
\begin{equation}
 \Delta\vv = - \frac{\dt/\ts}{(1+\dt/ts)}\, (\vv-\vg(t+\dt))  ~.
 \label{eq:pebble_deltav}
\end{equation}

In the limit $\dt/\ts \gg 1$, the velocity ends up very close to the gas velocity (as it should), but without the need to take very short time steps, resulting in a much more efficient time integration. In the limit $\dt/\ts \ll 1$, one recovers the normal kick-drift-kick behaviour.

We thus advance the velocity in the first (kick) step with:
\begin{equation}
 \vv_1 = \vv_0 + \frac{\dt}{(1+\dt/\ts)} 
 \left(\ag(\rr_0,t_0)+\ad(\rr_0+\vv_0\dt,t_0+\dt)\right) .
 \label{eq:kdk_step1}
\end{equation}
The particle position and time are then updated (drift step) with:
\begin{equation}
\rr_2 = \rr_0 + \vv_1 2 \dt  ~,
\label{eq:kdk_step2_pos}
\end{equation}
and
\begin{equation}
t_2 = t_0 + 2 \dt  ~.
\label{eq:kdk_step2_time}
\end{equation}
Finally, the third step (kick) advances the velocity using the acceleration at the end of the full time step, with
\begin{equation}
\vv_2 = \vv_1 + \frac{\dt}{(1+\dt/\ts)}
\left[\ag(\rr_2,t_2)+\ad(t_2)\right] ~.
\label{eq:kdk_step3}
\end{equation}

We have tested this formulation on particle drift in disks with known solutions \citep{Weidenschilling77}, recovering behaviour sufficiently accurate for our purposes for arbitrary values of Stokes numbers $St = \Omega \ts$.

The particle update procedure is computationally efficient, with the average update cost per cell on an Intel Ivy Bridge core increasing from 0.65 $\rm{\mu}$s without particles to 0.72 $\rm{\mu}$s with one particle per cell, indicating an average update cost of about 70 nanoseconds per particle.  Most of that time is spend looking up properties needed to calculate the particle-gas drag force.

\begin{figure*}
  \centering
  \subfigure[Midplane.]{\includegraphics[width=0.33\linewidth]{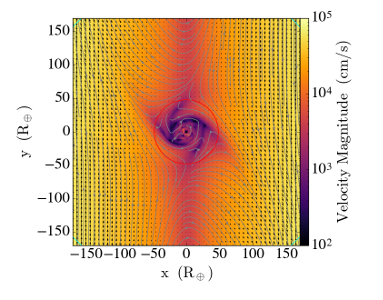}}\hfill
  \subfigure[$z$ = 10 $\rp$.]{\includegraphics[width=0.33\linewidth]{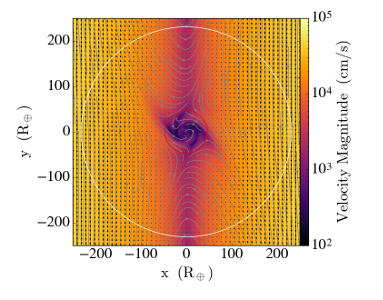}}\hfill
  \subfigure[$z$ = 25 $\rp$.]{\includegraphics[width=0.33\linewidth]{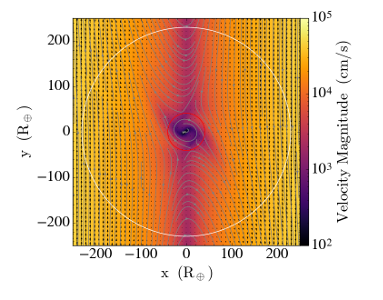}}\\
  \subfigure[$z$ = 75 $\rp$.]{\includegraphics[width=0.33\linewidth]{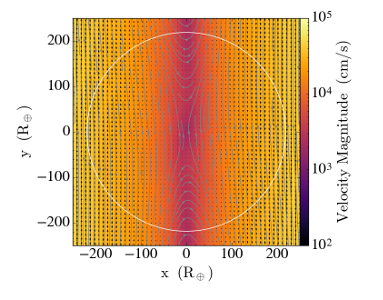}}\hfill
  \subfigure[Vertical component of the vorticity in the midplane.]{\includegraphics[width=0.33\linewidth]{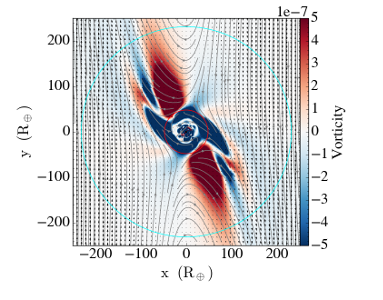}}\hfill
  \subfigure[$xz$ slice of $u_z$ at $y$ = 0.]{\includegraphics[width=0.33\linewidth]{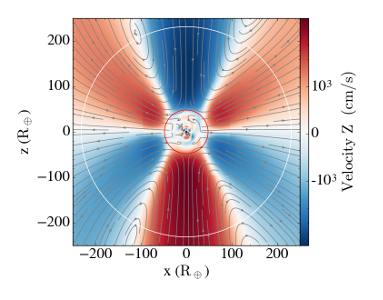}}
  \caption{Gas flows around the $M$ = 0.95 \ME\, embryo in the \texttt{m095t00} run. Panels (a-d) show the velocity magnitude at different heights above the midplane. The scale of the colour bar is deliberately held constant. Panel (e) shows the vertical component of the vorticity in the midplane and panel (f) shows $u_z$ in a $xz$ slice at $y$ = 0. Velocity streamlines are also shown.}
  \label{fig:gas_high_mass}
\end{figure*}

\subsection{Injection of pebbles}
\label{subsec:pebble_injection}
Once the simulations are considered to be fully relaxed, pebbles are injected into the system. The probability that a particular cell in a particular patch will spawn a macro-particle is proportional to $N_\mathrm{ppc}$, the number of particles per cell (which need not be an integer).  We set $N_\mathrm{ppc}$ to 0.05 in most runs, but increase it to 1.0 in a few runs to improve the statistics when studying the distribution of particles in small volumes.

The mass carried by a macro-particle is encoded in a `weight factor', $w = \rho \Delta V/N_\mathrm{ppc}$, where $\Delta V$ is the cell volume. The weight factor is thus proportional to the mass of gas in the cell where the macro-particle originates divided by $N_{ppc}$. An advantage of this approach is that the sum over all particles (or a subset defined by particle size or initial particle location) results in the total mass of those particles (to be renormalized by a preferred initial dust-to-gas ratio).

However, because the mass contained in any given cell can vary substantially, particles may need to be split to avoid the situation where the particle mass in a patch is represented by too few particles. For purposes of efficiency, the need for splitting is only checked when a particle passes from one patch to another. If an inbound particle carries more weight than, e.g., five times the initial average in the patch, it is split into a number of ``child particles''. The child particles continue to carry, also after any additional splits, a memory of their initial identity, which is needed for identifying sub-populations based on initial physical location.

When a macro-particle is created, it is assigned a random position inside a cell and a randomly chosen size:
\begin{equation}
s = 10^{-3\cal{R}}~\mbox{cm},
\label{eq:initial_size}
\end{equation}
where $\cal{R} \subseteq$ [0,1] is a random number. The spawned macro-particles are initially given the gas velocity of the cell. We spawn $\approx$ 2.6 million macro-particles in most runs, with 53 million in a few runs with $N_\mathrm{ppc} = 1$. We save a complete set of particle data when a snapshot of the gas dynamics is taken, but, for visualization purposes, we also separately save trajectory information for a smaller subset of macro-particles ($\sim$15 000) at every particle time step. These `tracer particles' are used to accurately track the trajectories of accreting particles and correspondingly to show the escape routes of particles that enter the Hill sphere and then leave again.

\section{Results: gas flows}
\label{sec:results_gas}
We begin by investigating the gas flow patterns around the different mass cores. In the last several years, a number of authors have studied the hydrodynamics in the vicinity of embedded planetary embryos (e.g.\ \citealt{dangelo2013, Ormel2015b, Fung2015, Masset2016, Cimerman2017, Lambrechts2017, Xu2017}, etc.) across a broad range of different conditions (e.g.\ different core masses, orbital distances from the parent star, disk structures), physical processes (isothermal/adiabatic/realistic EOS, with/without radiative energy transfer, magnetic fields, etc.). As illustrated below, our simulations are consistent with these previous studies, while offering both a higher resolution and a larger domain around the embryo.

\begin{figure*}
  \centering
  \subfigure[Midplane.]{\includegraphics[width=0.33\linewidth]{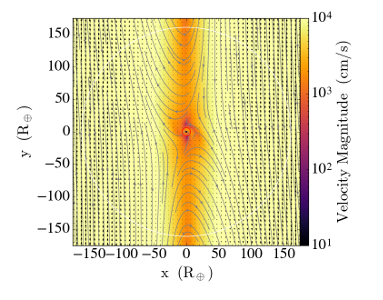}}\hfill
  \subfigure[$z$ = 10 $\rp$.]{\includegraphics[width=0.33\linewidth]{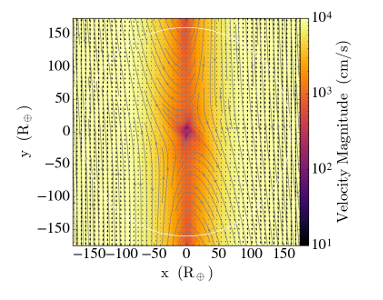}}\hfill
  \subfigure[$z$ = 25 $\rp$.]{\includegraphics[width=0.33\linewidth]{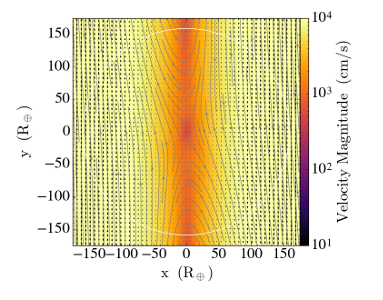}}\\
  \subfigure[$z$ = 75 $\rp$.]{\includegraphics[width=0.33\linewidth]{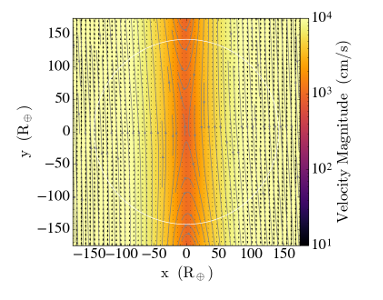}}\hfill
  \subfigure[Vertical component of the vorticity in the midplane.]{\includegraphics[width=0.33\linewidth]{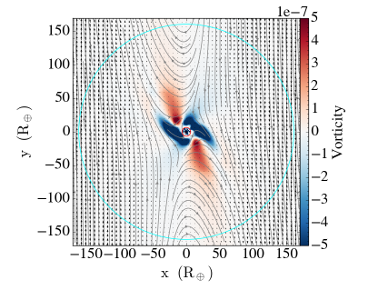}}\hfill
  \subfigure[$xz$ slice of $u_z$ at $y$ = 0.]{\includegraphics[width=0.33\linewidth]{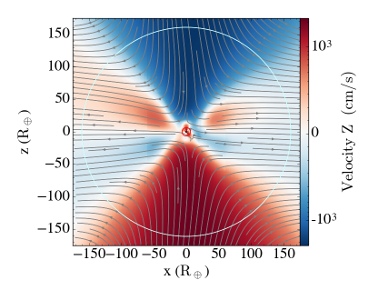}}\hfill
  \caption{Same as \Fig{gas_high_mass}, but for the M=0.096 M$_\oplus$ embryo in the \texttt{m01t00} run.}
  \label{fig:gas_low_mass}
\end{figure*}
\begin{figure*}
  \centering
  \subfigure[\texttt{m01t00} run.]{\includegraphics[width=0.33\linewidth]{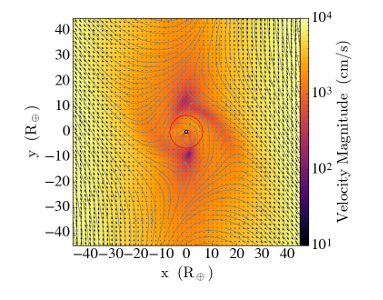}}\hfill
  \subfigure[\texttt{m05t00} run.]{\includegraphics[width=0.33\linewidth]{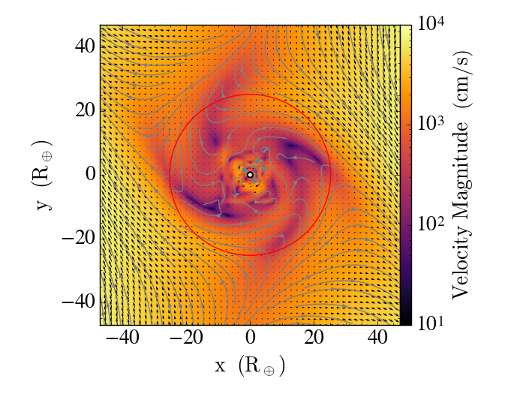}}\hfill
  \subfigure[\texttt{m095t00} run.]{\includegraphics[width=0.33\linewidth]{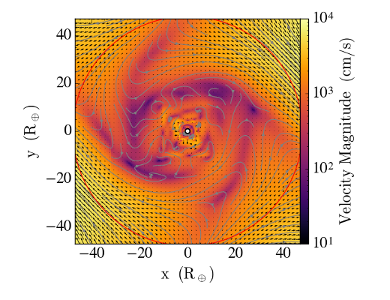}}\hfill
  \caption{Velocity magnitude in the midplane in the vicinity of the embryos as a function of embryo mass.}
  \label{fig:vel_mag_close}
\end{figure*}

\Fig{gas_high_mass} shows the details of the gas flow in the \texttt{m095t00} run for the $M$ = 0.95 \ME\ planetary embryo. For clarity, we zoom-in to the Hill radius (234 embryo radii, white circle). The red circle denotes the nominal Bondi radius, $R_{\rm B}$, which is $\approx 50 \rp$. The embryo is marked by a black circle, but is too small to be visible in these figures. The first four panels show the velocity magnitude in the disk midplane ($z$=0), and at $z$ = 10, 25, and 75 \rev{$\rp$}. As expected, the perturbation of the gas due to the embryo decreases with increasing height above the midplane, as indicated by the streamlines\footnote{Note, however, that the streamlines does not give an entirely accurate impression of the flow pattern because they only consider the velocity in the plane of the slice.}. The velocity magnitude \rev{in the horizontal plane 75 planet radii above the midplane} is nearly uniform along the $y$-direction, indicating that\rev{, at that height,} there are no substantial perturbations where the embryo is located, consistent with \citet{Ormel2015b}. We also observe that the width of the horseshoe orbits decrease with increasing height above the midplane, which agrees with the results of \citet{Fung2015}, but contradicts the isothermal results of \citet{Masset2016}. This might be attributed to the fact that we use a non-isothermal EOS, but further analysis of the horseshoe orbits is beyond the scope of this work. Panel (e) of \Fig{gas_high_mass} shows the vertical component of the vorticity measured relative to the rotating coordinate system. A clear vorticity pattern can be seen, with separators splitting the flow into regions that pass the embryo on the left and right sides. The flows passing on the ``wrong'' side contribute directly to positive (prograde, counter-clockwise) vorticity, which adds to the rotation of the flow.

Panel (f) of \Fig{gas_high_mass} shows the vertical velocity, $v_z$, in an $xz$ plane at $y$ = 0. A vertical gas flow pattern can be distinguished: gas flows towards the embryo from the poles and moves away at the midplane. Zonal flow patterns are visible as closed streamline loops. Vertical velocities inside R$_{\rm B}$ are of the order of a few cm s$^{-1}$, which agrees well with previous works(e.g.\ \citealt{Ormel2015b, Lambrechts2017}).

Qualitatively, the $M$ = 0.5 \ME\, case is very similar, thus the figure panels for this mass are presented in Appendix \ref{app:additional_figures}; \Fig{gas_medium_mass}.

\rev{The gas flows around the $M$ = 0.096 \ME\, embryo (\texttt{m01t00} run) behave somewhat differently, as illustrated in \Fig{gas_low_mass}. Because of the lower embryo mass, the perturbations away from pure shear flow are smaller, and horseshoe orbits therefore reach further in towards the embryo surface. The strong features in vorticity are correspondingly smaller. The vertical flows of the gas, meanwhile, approach the embryo over a broader extent radially, relative to the size of the Hill sphere. In our simulations, the gas flow patterns in the innermost region change with time, alternating between dominating gas inflows from the leading/outer horseshoe flow and the trailing/inner one. The locations of the stagnation points also vary with time. However, it is possible that part of this behaviour is related to the recursive refinement of the grid. We performed exploratory runs with the same mesh refinement, but without the planet embryo, and found root-mean-square velocities in the innermost region on the order of 10 m s$^{-1}$. This is similar to the case with the low-mass embryo present, and hence the small scale, low amplitude motions are unreliable. These motions are not caused by noise generated directly by the mesh refinement, but rather by slight differences in the cancellation of the force of gravity and pressure gradients, where small differences due to numerical resolution result in imperfect hydrostatic balance.}

\subsection{Gas dynamics close to the planetary embryos}
\label{subsec:dynamics_close}
One reason why this work stands out relative to previous studies is that we are able to resolve the detailed gas dynamics (and hence accurate particle paths) in the close vicinity of the planetary embryos. \Fig{vel_mag_close} shows the gas velocity magnitude in the vicinity of $R_\mathrm{B}$ for the $M$ = 0.096 \ME\, (panel (a)), $M$ = 0.5 \ME\, (panel (b)) and $M$ = 0.95 \ME\, (panel (c)) embryo masses. The $M$ = 0.096 \ME\, core is not massive enough to form a hydrostatic atmosphere -- the horseshoe flows constantly flush the atmosphere and only a tiny part of it ($\approx$ 2 $\rp$ from the embryo) is nearly static. With increasing embryo mass, more gas becomes bound and the velocity magnitude decreases. Nevertheless, there are always slow gas movements inside $R_\mathrm{B}$ due to both gas from the disk that reaches deeper interior layers and from small scale turbulence. Therefore, the primordial atmospheres are never fully isolated and always interact with the surrounding disk.
\begin{figure}
  \centering
  \includegraphics[width=1.0\columnwidth]{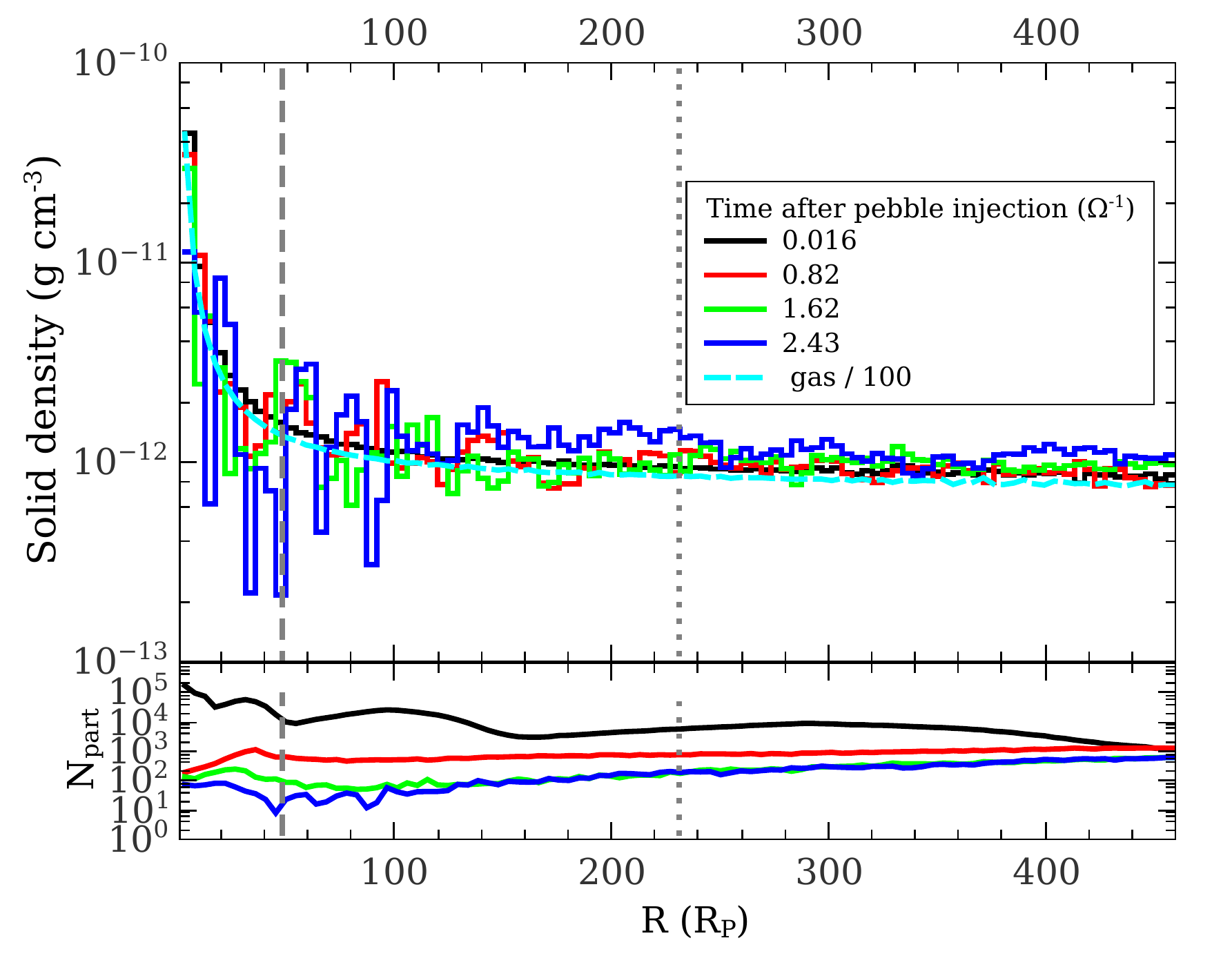}
  \caption{\rev{Mass} distribution \rev{of solids in radial shells} close to the embryo at different times for the \texttt{m095t10} run. The vertical dashed and dotted lines show the Bondi and Hill radii, respectively. \rev{The cyan line denotes the gas density in each shell, divided by a canonical dust-to-gas ratio of 100.} The lower panel shows the number of macro-particles as a function of radius. \rev{The bin width $dr$ is $4.66 \rp$.} As initially local macro-particles with low weights are replaced by macro-particles with larger weights from larger distances, the statistical noise increases.
}
  \label{fig:m095t10_peb_shell}
\end{figure}

\section{Results: particles}
\label{sec:results_particles}
An obvious advantage of our large simulation box is the large particle reservoir. Smaller simulation boxes, especially with non-periodic $x$-boundaries, often experience a depletion of macro-particles. This make\rev{s} the accurate evaluation of accretion rates more difficult, and additional injections of particles may be needed (e.g.\ \citealt{Xu2017}).

In \Fig{m095t10_peb_shell}, we show time-series of particle mass density from run \texttt{m095t10} across a range of integrated radial layers (shells) from the embryo surface to twice the Hill radius. \rev{B}lack denotes the first snapshot after the injection of the pebbles (after 40 time units) while the other lines are plotted at 4 000 time unit intervals (which corresponds to $\approx$0.8 of an orbital period). \rev{We also plot the integrated density of the gas divided by 100 (the canonical gas-to-dust ratio), in cyan for the same radial bins.} The bottom panel shows the number of macro-particles in each radial bin. Here, the initial dips on the black curves stem from changes in the level of refinement. As time goes on, the number of macro-particles decreases due to accretion and transport, but new particles are continuously coming from further out. These new particles originate in patches with lower numerical resolution (i.e.\ larger cell sizes), and thus carry larger representative weights. \rev{The increasing noise level, meanwhile, is due to the increasing contribution from these higher statistical weight macro-particles.}

\rev{Even after 2.4 orbits, \rev{\Fig{m095t10_peb_shell} shows} no indication of depletion of solids -- the mass influx through a given shell is sufficient to replace any accreted or lost particles. This implies that our initial injection of particles is sufficiently numerous to allow accurate measurements of accretion rates without further injections.} Similar trends are seen for the other embryo masses; as the solid density is proportional to the gas density, we choose to not show separate figures for them.

\begin{figure}
\centering
    \includegraphics[width=\columnwidth]{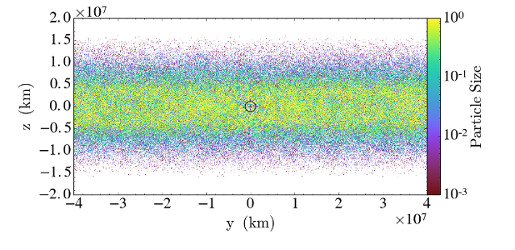}
    \caption{Vertical distribution of pebble sizes in the \texttt{m095t00} run 1.6 orbits after the injection of macro-particles.}
    \label{fig:pebbles_m095t00d10_yz}
\end{figure}

\begin{figure*}
  \centering
  \subfigure[Particle distribution over a 2R$_{\rm H}$ vertical extent above/below the midplane.]{\includegraphics[width=0.66\columnwidth]{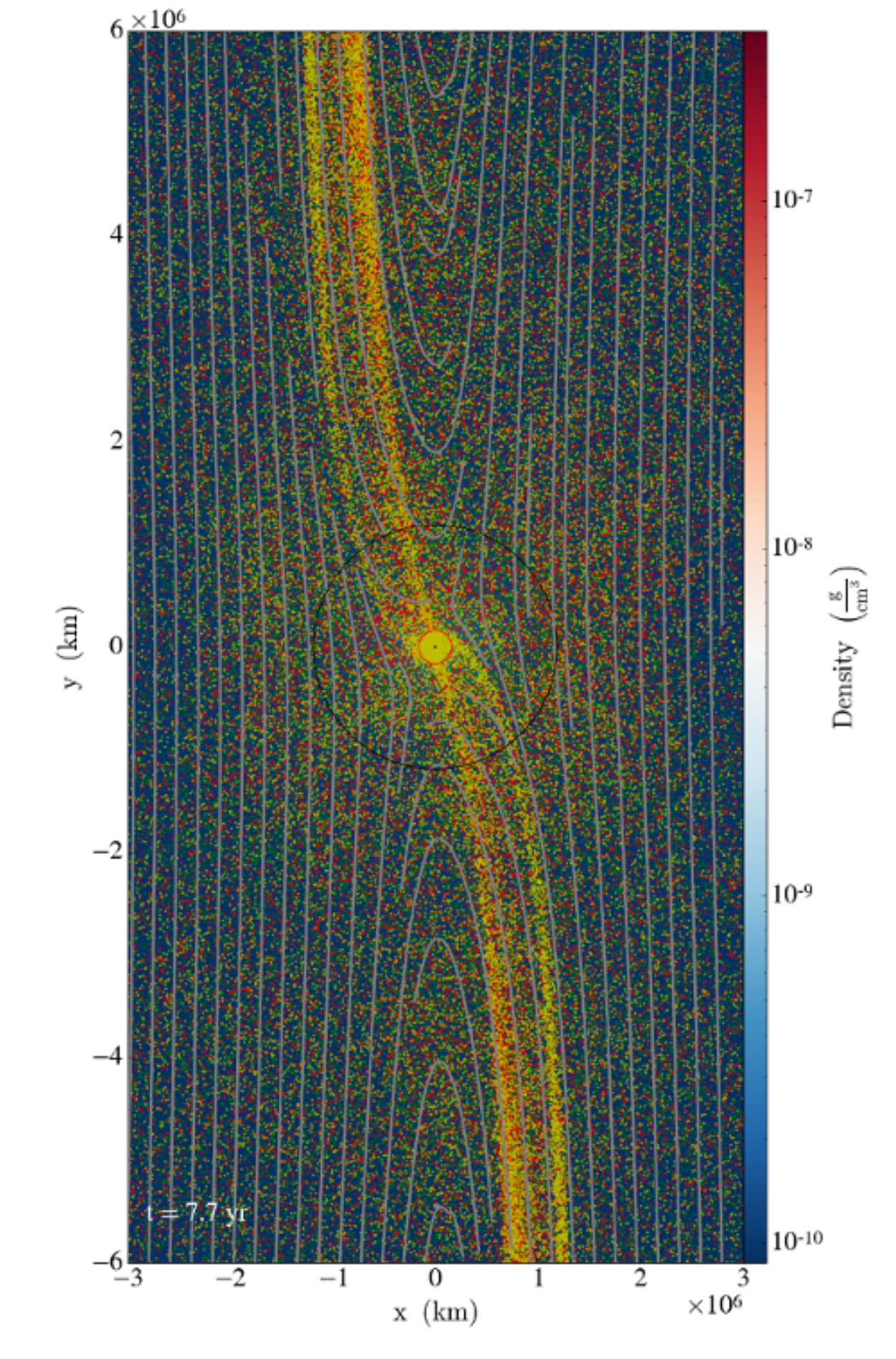}}\hfill
  \subfigure[Same as (a), but with a 2R$_{\rm B}$ vertical extent]{\includegraphics[width=0.66\columnwidth]{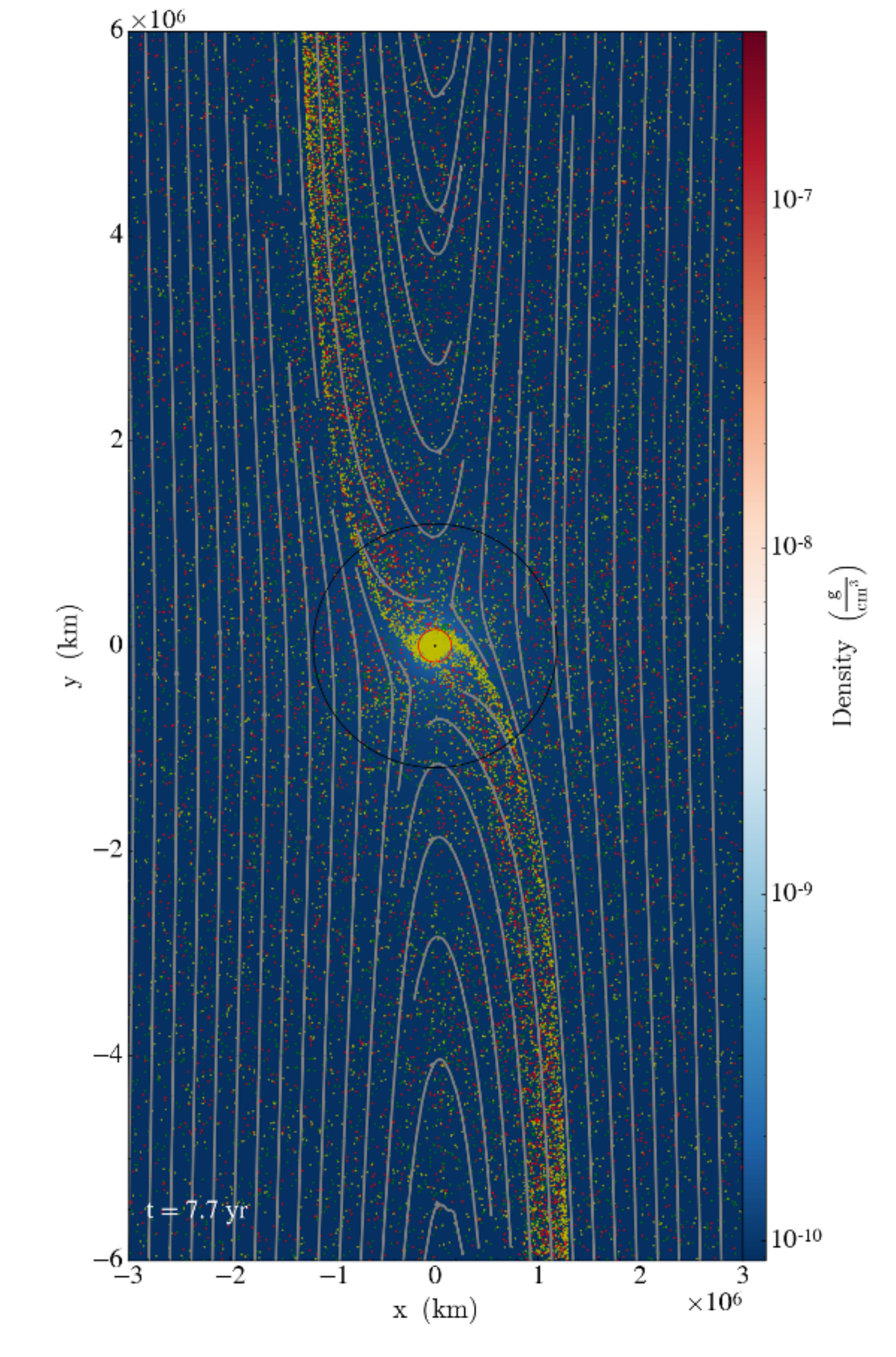}}\hfill
  \subfigure[A zoom-in of panel (b).]{\includegraphics[width=0.66\columnwidth]{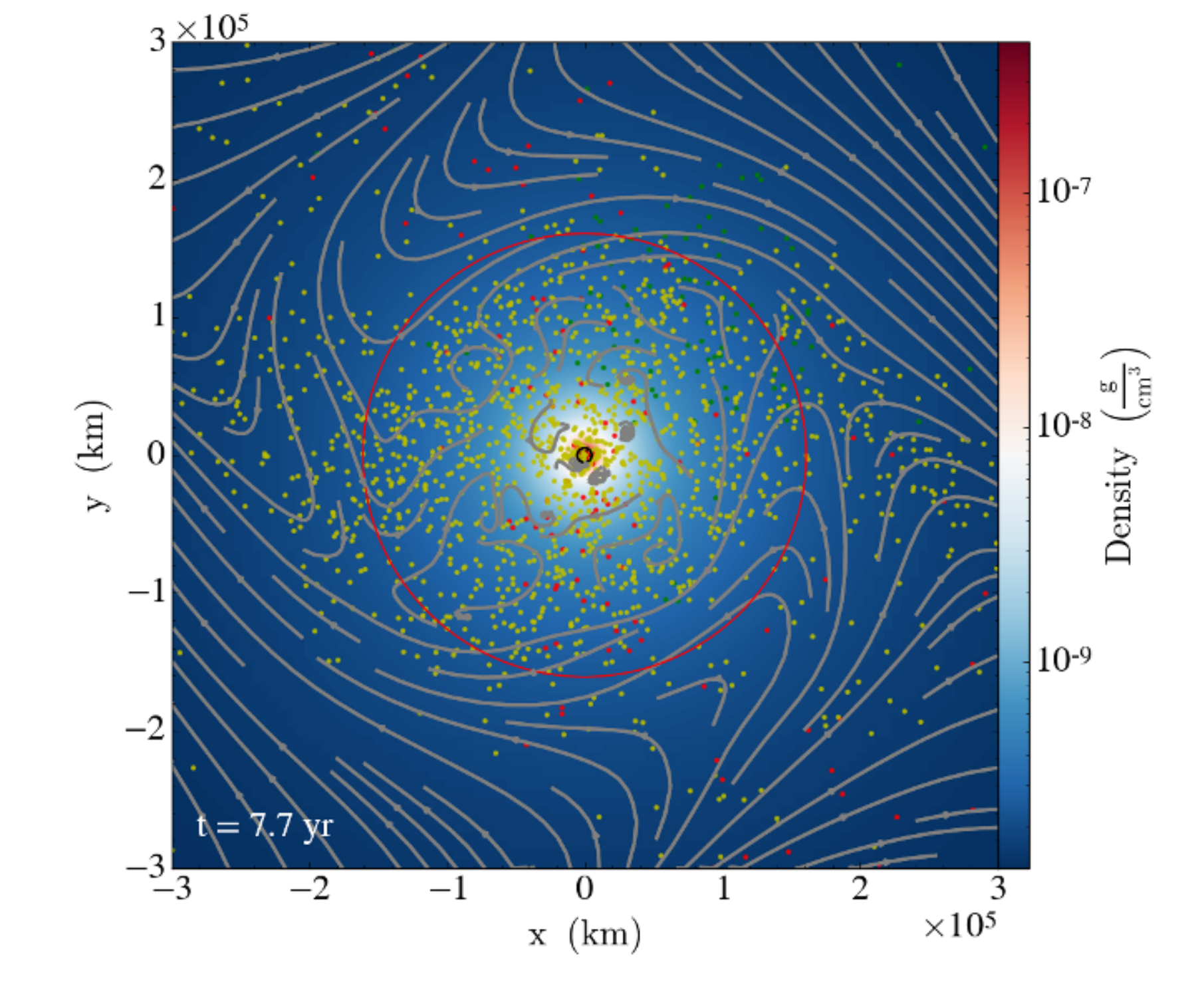}}\hfill
  \caption{Gas density and \rev{spatial distribution of particles} close to the midplane in the \texttt{m095t00} run. The \rev{particles} here are split into 3 size bins: $0.001 < s \leq 0.01$ cm (yellow), $0.01 < s \leq 0.1$ cm (red), and $0.1 < s \leq 1$ cm (green). \rev{The black circle denotes the Hill sphere, while the red circle denotes the Bondi radius.}}
  \label{fig:pebbles_m095t00d10_midplane}
\end{figure*}

\subsection{The distribution of particles}
\label{subsec:particle_distribution}
We initialize particles uniformly in the dust-to-gas ratio, and we therefore assume the initial vertical stratification of particles is the same as that of the gas. Particles then settle towards the midplane due to the vertical force of gravity, with settling time scales of the order of $1/(\ts \OmegaK^2)$. \rev{The} settling time scale \rev{for small and intermediate size particles} particles is long relative to the duration of our experiments, while it is comparable to our run times for the largest particles. Lacking proper turbulence in the disk, the settling of particles is anyway not very realistic---turbulence stirs up the particles and controls their scale height. Additionally, zonal gas flows (e.g.\ panel (f) of \Fig{gas_high_mass}) close to the Hill sphere are capable of stirring up well-coupled particles.

Therefore, instead of working with ill-defined and unknown equilibrium populations, we choose instead well-defined, localized sub-populations with initially constant dust-to-gas ratios for our measurements of accretion rates. Tests have shown that it takes only $\sim$0.5 orbits after the injection of particles for the measured accretion rates to stabilize. The hydrodynamic flow patterns are established quickly as well, taking approximately the same amount of time. This shows that our initial relaxation of the flow patterns, which was allowed to continue for $\sim$6 orbital periods, was certainly sufficient and allows for accurate measurements of accretion rates.

\begin{figure*}
  \centering
  \subfigure[\texttt{m095t10}]{\includegraphics[width=0.65\columnwidth]{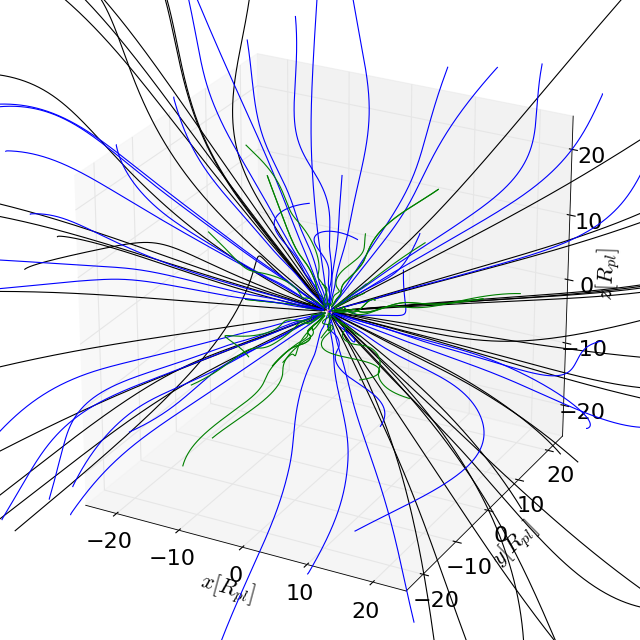}}\hfill
   \subfigure[\texttt{m05t10}]{\includegraphics[width=0.65\columnwidth]{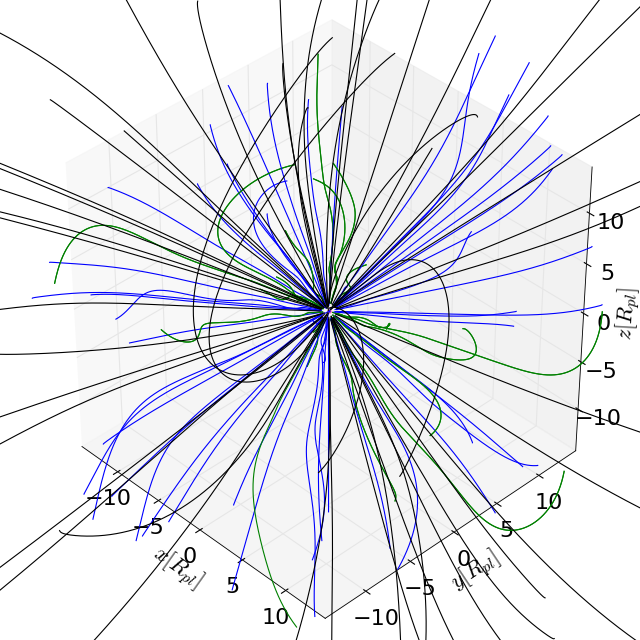}}\hfill
  \subfigure[\texttt{m01t10}]{\includegraphics[width=0.65\columnwidth]{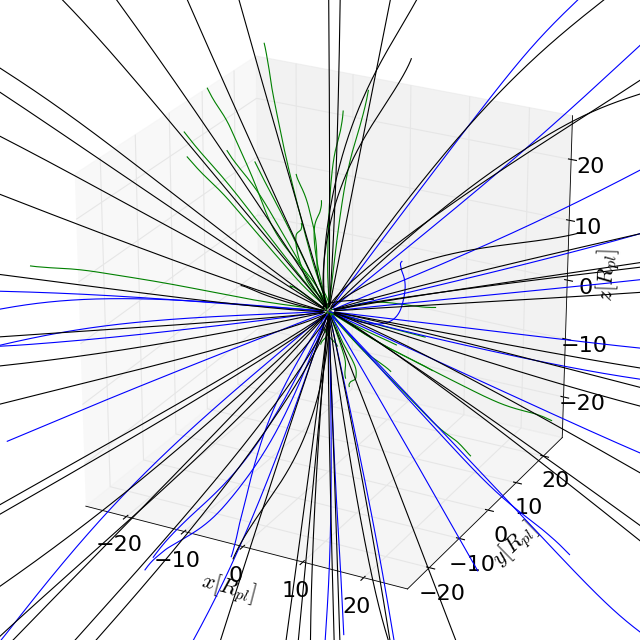}}\hfill
  \caption{Trajectories of accreted particles. The axes are given in units of embryo radii. Green, blue and black paths are for particle sizes $0.001 \leq s < 0.01$  cm, $0.01 \leq s < 0.1$ cm, and $0.1 \leq s < 1$ cm, respectively. }
  \label{fig:pebble_trace}
\end{figure*}
\begin{figure}
  \centering
  \subfigure[\rev{Spatial distribution of particles} over a $2R_{\rm B}$ vertical extent above/below the midplane. The black circle denotes the Hill sphere.]{\includegraphics[width=0.66\columnwidth]{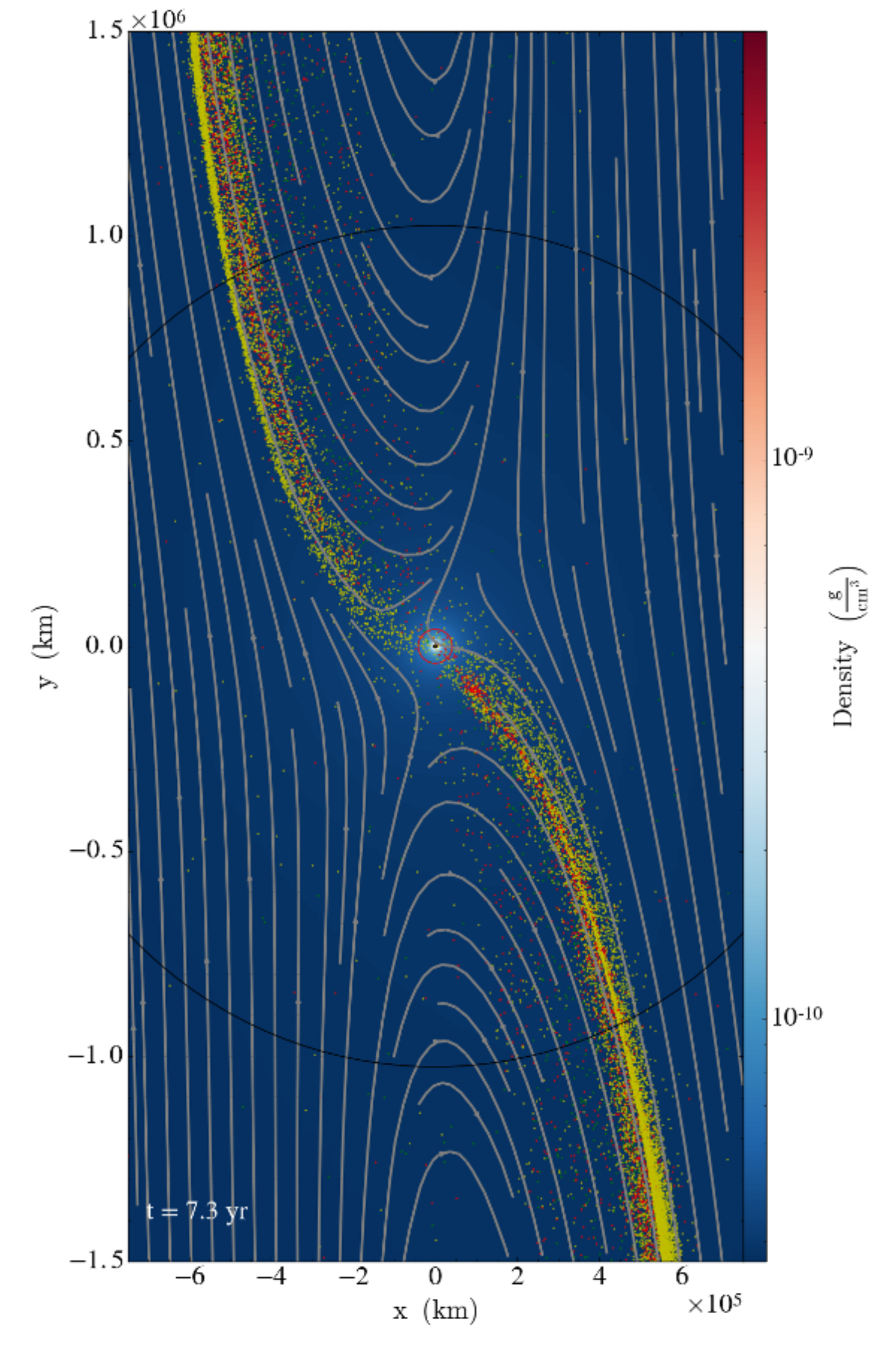}}\\
  \subfigure[A zoom-in of panel (a) to the Bondi sphere. The red circle denotes the Bondi sphere.]{\includegraphics[width=0.66\columnwidth]{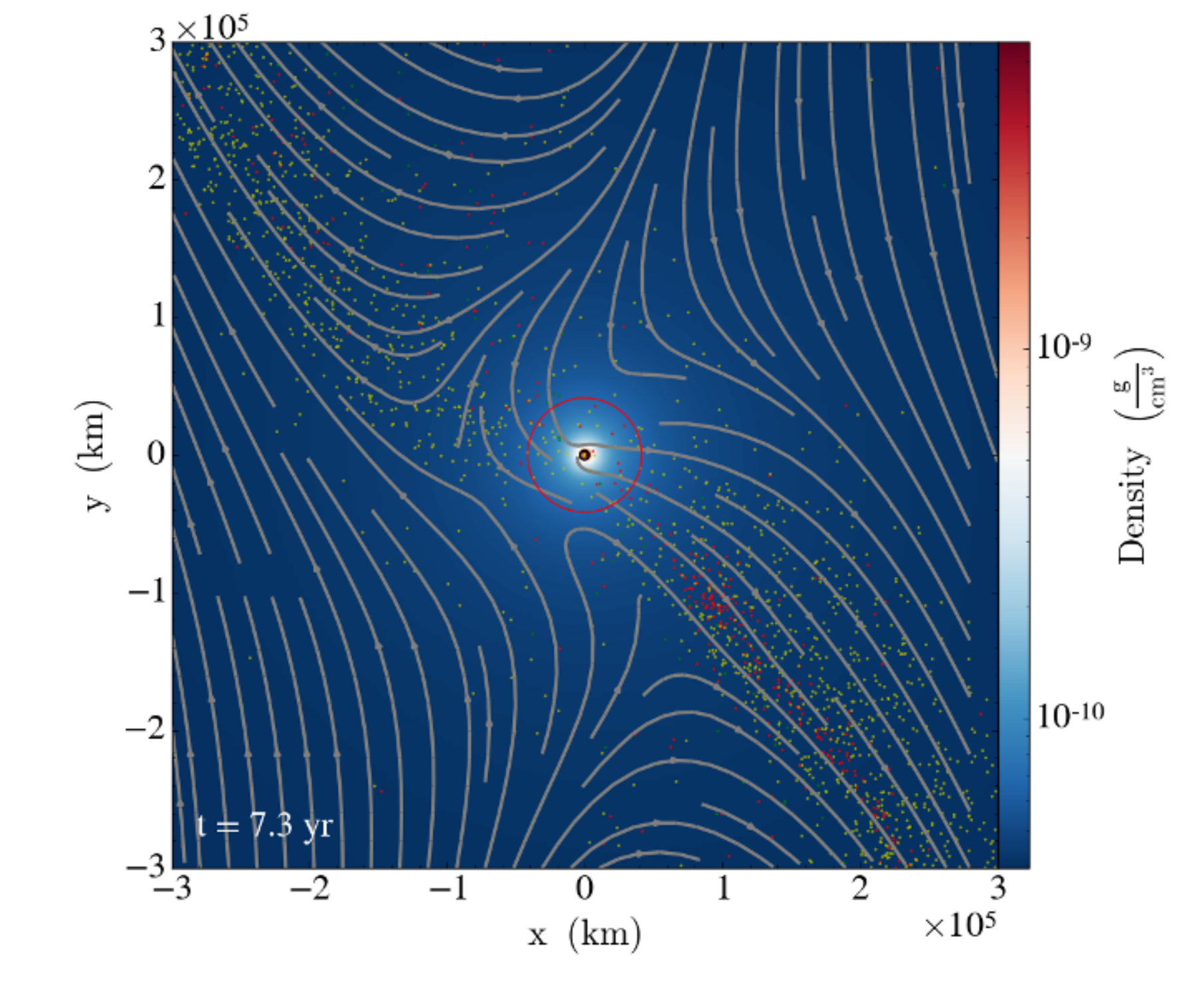}}
  \caption{Gas density and \rev{spatial} distribution \rev{of particles} close to the midplane in the \texttt{m01t00} run. The pebbles here are split into 3 size bins: yellow $s \leq 0.01$ cm, red $0.01 < s \leq 0.1$cm and $s > 0.1$ cm are green.}
  \label{fig:pebbles_m01t00d11_midplane}
\end{figure}

\Fig{pebbles_m095t00d10_yz} shows the vertical distribution of pebbles in the \texttt{m095t00} run, 1.6 orbits after the injection of macro-particles. The smallest particles ($s \leq 0.01$ cm; dark colours) continue to show essentially the vertical distribution of the gas, while larger particles ($s > 0.1$ cm; green to yellow) have begun to settle. This figure also nicely illustrates how small a perturbation the Hill sphere is relative to the size of our computational domain.

\Fig{pebbles_m095t00d10_midplane} shows the \rev{spatial} distribution of particles and their sizes at the midplane, two orbits after the injection of particles. This time, the particles are split into 3 size bins: $s \leq 0.01$ cm (yellow), $ 0.01\leq s<0.1$ (red) and $s > 0.1$ cm (green). The vertical extent \rev{of} particles \rev{we display} is limited to $1\RH$ in panel (a) and $2\RB$ in panels (b,c), above and below the midplane. One instantly notices the two leading and trailing arms, stretching away from the embryo. The large concentration of particles there indicates that these particles are moving away from the embryo very slowly. The wakes are located just inside the separatrix sheets of the horseshoe orbit, with the smallest particles mostly concentrating in the outermost regions and the larger ones \rev{lying} closer to the \rev{orbit of the embryo}. These macro-particles carry a relatively small statistical weight\rev{, because they originate at patches with high resolution and small mass per cell}, so the features are somewhat artificially enhanced in the \rev{F}igure, but \rev{it} therefore also nicely illustrates the preferred `leaving' paths of the particles. A large fraction of these particles enter \rev{but then} later \rev{leave} the Hill sphere, and the features are therefore concentrated in the midplane. 

A zoom-in to $\RB$ (panel (c)) reveals that there is a concentration of small particles ($s < 0.1$ cm) inside the Bondi sphere. Larger particles do not linger there for long since they are effectively accreted. The figure shows particle paths based on high-cadence positions of tracer-particles, sub-divided into 3 size bins. The green colour represents $0.001\leq s < 0.01$ cm, blue $ 0.01\leq s<0.1$ cm and black $s 0.1\leq s< 1$ cm size particles. The smallest particles spend around 0.1$\Omega^{-1}$ inside the Bondi sphere before they reach the embryo. In contrast, particles with $s \geq 0.01$ cm, are accreted on average 10 times faster, after they reach the Bondi sphere. Alternatively, some of the smallest particles, if they linger far enough from the embryo, are later carried out by the horseshoe gas flows.

In these runs we do not consider convection driven by the accretion heating (but see below), particle growth and evaporation. Sufficiently strong convection could possibly prevent the smallest particles from reaching the core. Also, as the gas temperature close to the embryo ($\approx$ 4500 K near the $M$ = 0.95 \ME\, embryo) greatly exceeds the melting point of any solid, evaporation of solids and the resulting enrichment of the atmosphere with heavy elements (e.g. \citealt{alibert_maximum_2017,brouwers_how_2017}), is bound to play an important role in models with a realistic treatment of evaporation. As shown by \Fig{eos}, a more realistic equation of state leads to lower temperatures at the base of the atmospheres, but still large enough to cause evaporation.

\Fig{pebbles_m01t00d11_midplane} similarly shows the \rev{spatial} distribution of pebbles close to the midplane (inside $2\RB$) in the \texttt{m01t00} run. It shows a very similar picture --- escaping particles are trapped in the leading/inner and trailing/outer horseshoe orbit parts. The zoom-in of the \rev{F}igure (panel b) reveals an interesting, but expected difference --- there is no accumulation of particles inside the Bondi sphere. There are two \rev{factors} that contribute to \rev{this}: the gas flows may be strong enough to readily carry small and intermediate size particles away, and larger particles are rapidly accreted. Panel (c) in \Fig{pebble_trace} reveals the answer: the stopping time, even of the smallest size particles we consider in our runs, is short enough that the gravitation of the embryo is not damped by the gas drag enough to prohibit particles from being accreted. The smallest particles in our simulation take merely a 0.04 $\Omega^{-1}$ to be accreted. And the number of these smallest particles crossing the Bondi sphere is much larger, compared to the $M$ = 0.95 \ME\, embryo. Here we remind the reader that we do not trace all the particles with this high cadence, therefore only a fraction of all the accreted particles are shown in \Fig{pebble_trace} (we also, for greater clarity, limit the number of  traces shown in these figures to 150).

\Fig{pebble_trace} shows evidence for a variation of hydrodynamical deflection (e.g. \citealt{Sellentin2013}) - higher mass embryos are accompanied by much denser primordial atmospheres, in which the stopping times become smaller. Whereas the majority of the particles in the distribution we consider here have nearly ballistic orbits when being accreted by the lowest mass embryo (\Fig{pebble_trace}, panel (c)), they become significantly affected by the higher gas density envelopes in panels (b) and (a). In the latter, the highest-mass embryo case, the motion of the gas almost completely controls the motion of particles with size $s < 0.01$ cm.

\begin{figure*}
  \centering
  \includegraphics[width=0.33\textwidth]{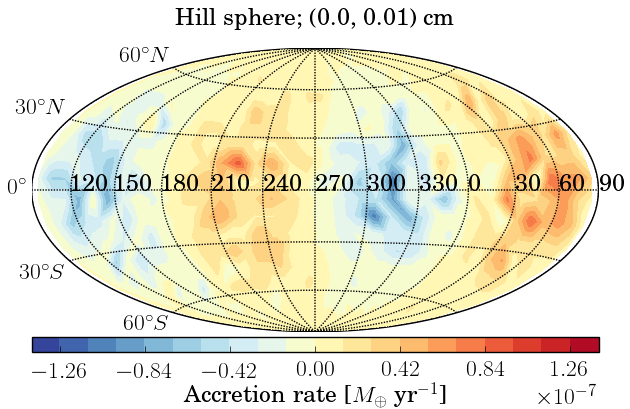}\includegraphics[width=0.33\textwidth]{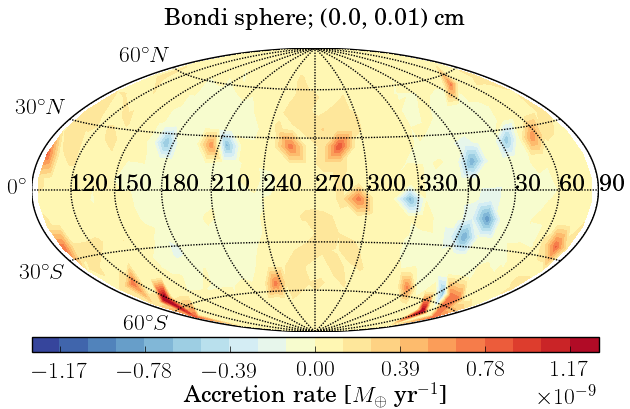}\includegraphics[width=0.33\textwidth]{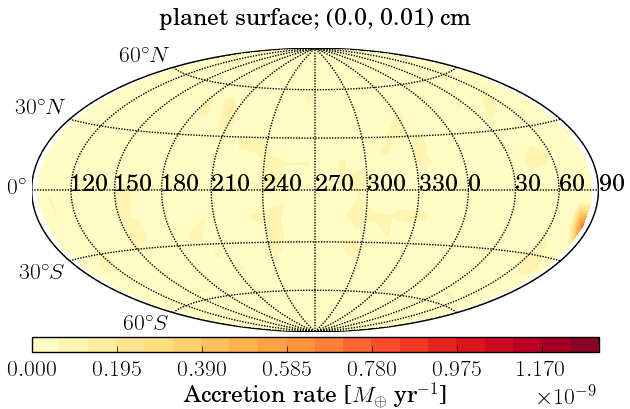}
  \includegraphics[width=0.33\textwidth]{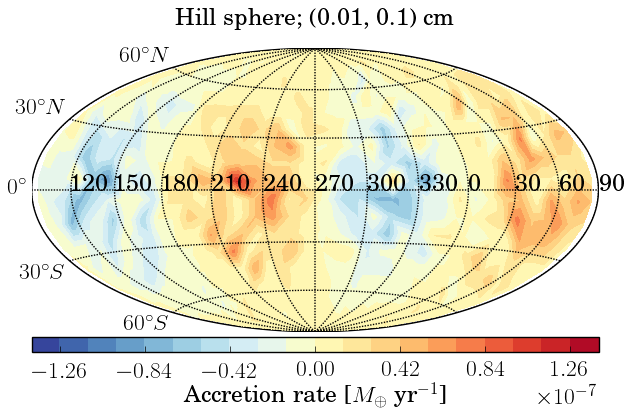}\includegraphics[width=0.33\textwidth]{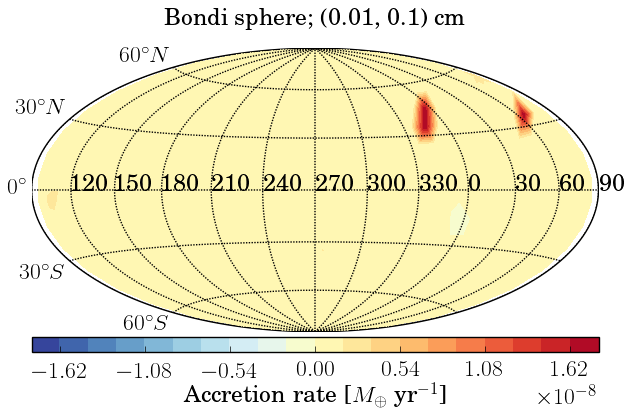}\includegraphics[width=0.33\textwidth]{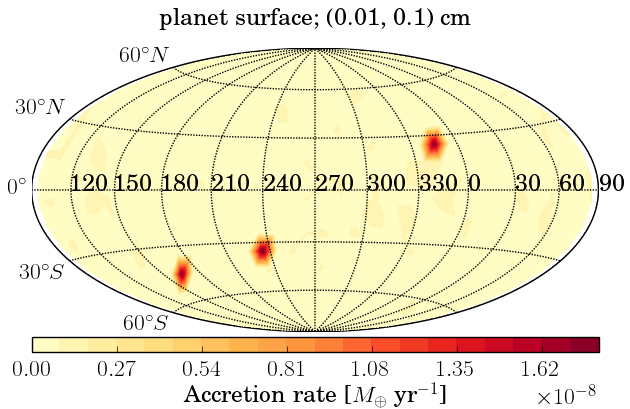}
  \includegraphics[width=0.33\textwidth]{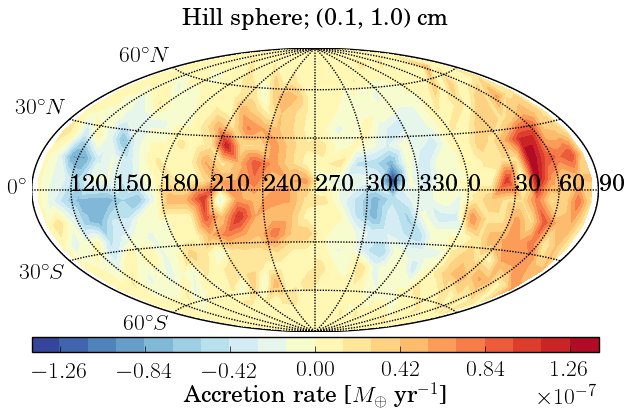}\includegraphics[width=0.33\textwidth]{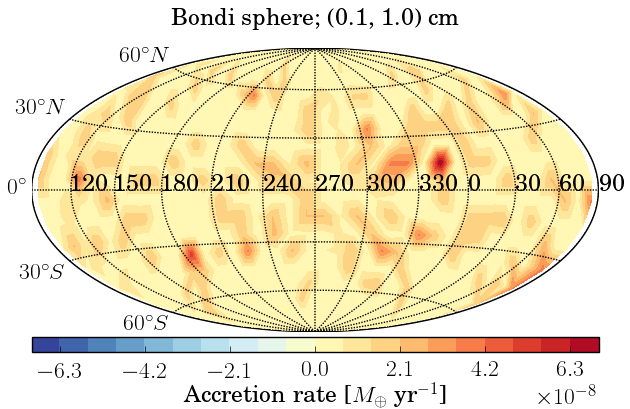}\includegraphics[width=0.33\textwidth]{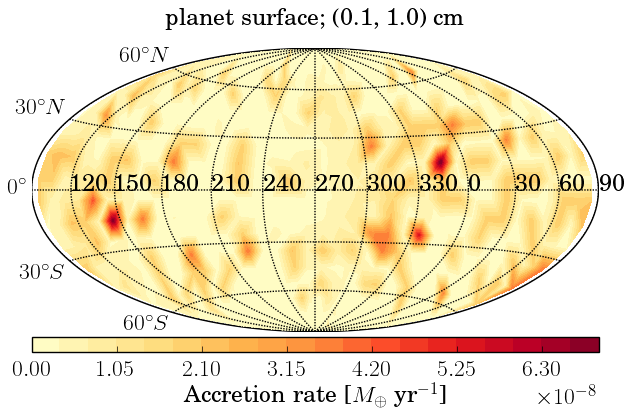}
  \caption{From left to right, the mass accretion rate across the Hill sphere, Bondi sphere and embryo surface, respectively, over the course of the \texttt{m095t10} run, displayed as a Hammer projection. From top to bottom are the particle accretion rates for three size bins $0.0 \leq s \leq 0.01$ cm, $0.01 < s \leq 0.1$ cm and $0.1 < s \leq 1.0$ cm. A longitude of 270$^\circ$ corresponds to the -$y$-direction and a latitude of 0$^\circ$ corresponds to the disc midplane.}
  \label{fig:accretion_m095t10d10p}
\end{figure*}

\subsection{Accretion of pebbles}
When pebbles approach an embryo they are subject to an increasing two-body force and start to veer towards the embryo. Although from \Fig{pebble_trace} it might seem that they come from all over the place, the figure greatly under-represents the particles (150 traces selected out of 15 000, again randomly selected from $\approx$2.6 million macro-particles), and the scales are concentrated to the closest 10-20 planetary radii around the embryos. The full ensemble of trace-particles in fact show distinct inflows and outflows through the Hill sphere. 

\begin{figure}
\centering
    \includegraphics[width=\columnwidth]{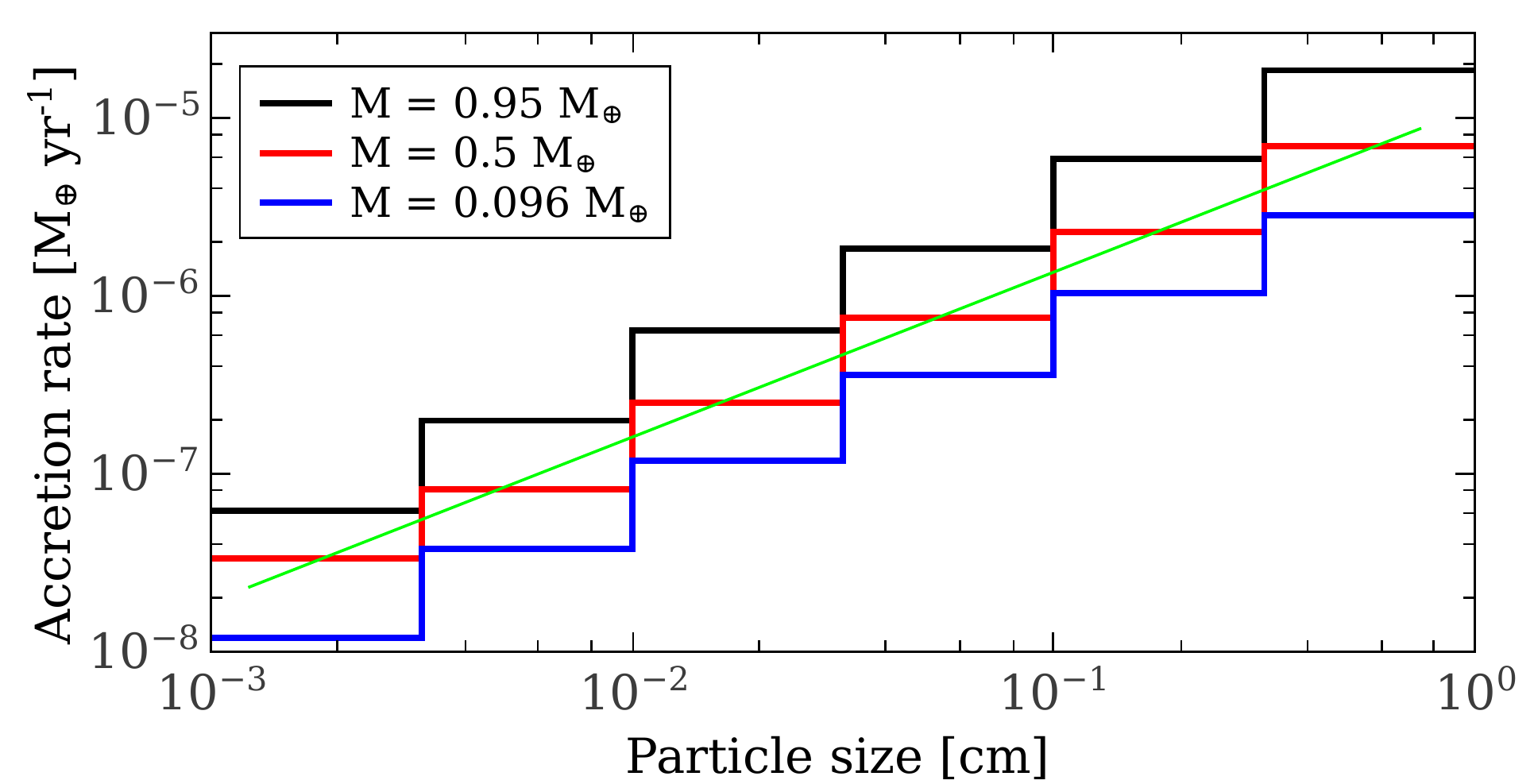}
    \caption{Pebble accretion rates \rev{onto the embryo}, for particle sizes within each size bin, drawn from an initial distribution consisting of particles within $\pm R_\mathrm{H}$ from the midplane, for the three embryo masses, 0.95 \ME (black), 0.5 \ME (red), and 0.096 \ME (blue). The green line shows a mass accretion rate proportional to particle size, $s$.}
    \label{fig:accretion_size_distribution}
\end{figure}

\begin{figure}
\centering
    \includegraphics[width=\columnwidth]{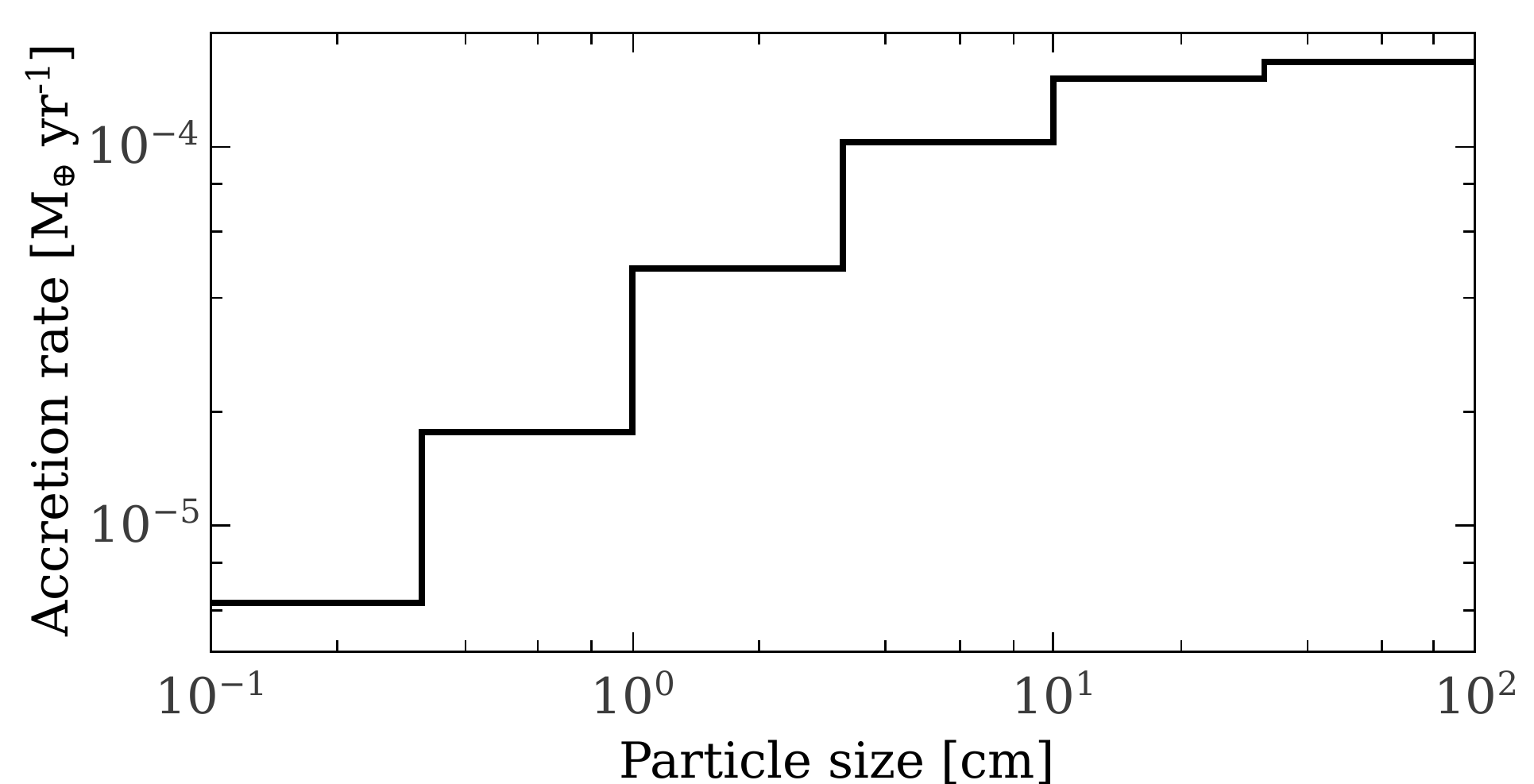}
    \caption{Pebble and boulder accretion rates \rev{onto the embryo}, for particle sizes between 0.1 - 100 cm, subdivided into 6 size bins, drawn from an initial distribution consisting of particles within $\pm R_\mathrm{H}$ of the midplane for the 0.95 \ME\, embryo mass \rev{case}.}
    \label{fig:acc_rate_boulder}
\end{figure}

\Fig{accretion_m095t10d10p} plots the mass accretion rates across the Hill sphere, Bondi sphere and the embryo surface over the course of the \texttt{m095t10} run, displayed as a set of Hammer projections. The macro-particles are sub-divided into 3 representative size-bins ($0.0 \leq s \leq 0.01$ cm, $0.01 < s \leq 0.1$ cm and $0.1 < s \leq 1.0$ cm). The figures show that the bulk of the net accretion comes from near the midplane, and predominantly from the large size particles. As they do not feel much gas drag, some of the largest particles can and do cross the Hill sphere from arbitrary directions. The dominating inflow regions are associated with the inner/trailing and outer/leading horseshoe orbits, and corresponding outflows in the outer/trailing and inner/leading horseshoe orbits. The accretion rates at distances $\RB$ are much more isotropic, without clearly preferred trajectories. The modest outflows are mainly close to the \rev{midplane region}. Not all \rev{of} the mass accreted through the \rev{Bondi} sphere reaches the embryo surface during the time scales we consider here. The missing mass is contained in the smallest particles, which trace the gas flows very closely. Having even the slightest turbulence in the primordial atmosphere (which is always the case, as discussed earlier) traps these particles and prohibits them from reaching the surface over sufficiently small time scales, but nonetheless, they eventually are accreted. \rev{That said, as the experiments approach a steady-state, the net accretion rate across a sphere of a given radius (e.g.\ $\RH$ or $\RB$) centred on the embryo approaches parity with the accretion rate onto the embryo.} The two top rows of \Fig{accretion_m095t10d10p} also show how inefficiently the subsets of smaller particles are accreted. Very similar trends are present in the other two embryo mass cases we consider in this work, therefore, the figures for $M$= 0.5 \ME\, and $M$ = 0.096 \ME, embryos are in  Appendix \ref{app:additional_figures}.

\revtwo{To better illustrate the dependence of accretion on particle size, an animation showing the trajectories of pebbles in the vicinity of an $M$ = 0.95 \ME\, embryo (\texttt{m095t10}) is available on-line as supplementary material (and on YouTube \footnote{\url{https://youtu.be/2GNgoc71qIw}}). In the animation, cyan represents $0.0 \leq s \leq 0.01$ cm particles, red represents $0.01 < s \leq 0.1$ cm particles and yellow represents $0.1 < s \leq 1.0$ cm particles. Velocity streamlines are shown in green. The coloured circles meanwhile denote the Hill radius (magenta) and the ``nominal'' effective accretion radii for unperturbed flow, $r_\mathrm{eff}$ (\citealt{Lambrechts2012}, eq.\ 40; see below) for the three, aforementioned size bins. The animation clearly illustrates that: (i) particles that pass within their respective $r_\mathrm{eff}$ are accreted; (ii) small, well-coupled particles are particularly affected by the gas flow and often do not reach $r_\mathrm{eff}$.}

\begin{figure}
\centering
    \includegraphics[width=\columnwidth]{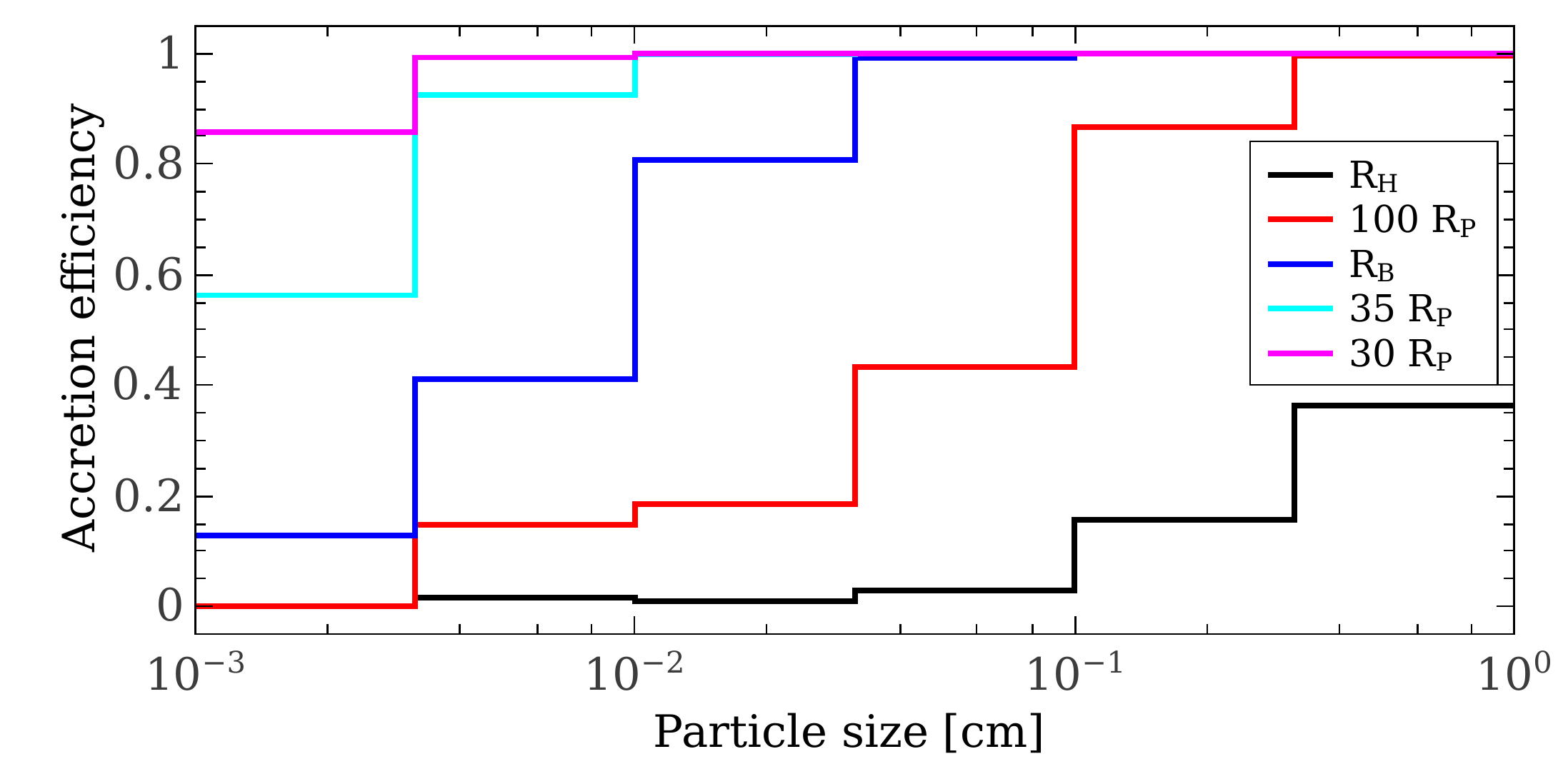}
    \caption{Local pebble accretion efficiency (\rev{the} net accretion rate divided by \rev{the} \rev{radially inward} mass flux across a sphere of a particular radius) as a function of particle size, \rev{for spheres of several different radii,} drawn from an initial distribution of particles within $\pm R_\mathrm{H}$ of the midplane for the \texttt{m095t10} run.}
    \label{fig:acc_eff_m095}
\end{figure}

\subsection{Quantitative measurements of accretion rates}
\label{subsec:quantitative_rates}
To allow selection of arbitrary sub-populations, our initial particle population is distributed all over space, with weights proportional to the local cell gas mass.  To obtain accretion rates pertaining to a disk where particles have settled towards the midplane, we select the sub-population that initially resides within one Hill radius from the midplane, and subsequently measure accretion rates based on only those particles, renormalising their macro-particle weights $w$ so they correspond to a total mass equal to the assumed dust-to-gas ratio (0.01) times the total gas mass in the experiment.

By sub-dividing the particle distribution into 6 size bins, we can study the dependency of the accretion rates on size. \Fig{accretion_size_distribution} shows that the net accretion rate onto the embryo scales linearly with the size $s$, which is different than the $s^{2/3}$ scaling predicted by e.g. \cite{Ormel2010}; \cite{Lambrechts2012}, assuming unperturbed shear flow conditions. As acknowledged in \citet{Lambrechts2012}, these conditions only hold when the mass of the planet embryo is very small. The key reason this scaling does not hold in our simulations is because the flow pattern (i.e.\ horseshoe orbits with stagnation points/separatrix surfaces) and density (with contrasts of the order of $10^3$) in the vicinity of the planet are very different from unperturbed shear flow. Small, well-coupled particles are most strongly affected by these differences; they tend to follow gas streamlines, and therefore follow horseshoe orbits or are deflected around the planet. As a result, the dependence of the accretion rate on particle size becomes steeper than $2/3$.  The measured linear scaling is valid even for larger particles, all the way to dm-sized particles, as illustrated in \Fig{acc_rate_boulder}. The accretion rate saturates (no longer increases with particle size) for particles larger than about 1 meter.

\rev{A crude measure for the importance of flow effects for pebble accretion can be obtained by equating the pebble accretion impact radius in the shear limit ($\reff \sim \tau^{1/3} \RH$, where $\tau$ is the Stokes number) with the Bondi radius $\RB$---the scale on which flow effects are expected to become important. For $\reff < \RB$ flow effects may be expected to change accretion rates that have been derived under the assumption of an unperturbed gas flow. Equating the radii, we find a threshold mass of $\Mps \sim h^3 \tau^{1/2} \Mstar$, where $h$ is the disk aspect ratio. Hence above $\Mps$ flow effects may modify pebble accretion estimates based on assumptions of pure shear flow. This expression also states that smaller particles (smaller $\tau$) are more severely affected, qualitatively consistent with what we find. However, to explain the reduction quantitatively, i.e., to explain the linear dependence of $\dot{\rm{M}}$ on $\tau$ that we find, requires a more thorough analysis, which will be considered in a future work.}

The runs with 56 million macro-particles give similar results as the corresponding runs with 2.6 million particles, with less than 10\% difference in the accretion rates through the Hill sphere and less than 1\% difference at the Bondi sphere. Thus the results are rather robust, even when using only 2.6 million macro-particles. \rev{Naturally, however, the greater the number of particles, the smaller a fraction of the population one can sample from and still retain} statistical significance; i.e., one has the ability to perform more focused studies of particle behaviour, as a function of size and initial position. The $M$ = 0.95 \ME\, case, for example, initially has only about 20\% of the macro-particles within one Hill radius of the midplane. Thus, to study midplane accretion rates it is useful to have a large particle population available.

An estimate of the maximum accretion rate may be obtained by taking the dust-to-gas ratio times the average gas mass flux through the Hill sphere. For our $M$ = 0.95 \ME\ embryo immersed in our 1/10 MMSN disk\rev{,} this flux is $\approx 4\ 10^{-3}$ \ME yr$^{-1}$. With a dust-to-gas ratio of 0.01, we thus estimate a maximum accretion rate of about $4\ 10^{-5}$ \ME yr$^{-1}$. This rate is exceeded by about a factor of five in the 0.3 -- 1 m size bin in \Fig{acc_rate_boulder}. By comparison, the `Hill rate'\rev{,} $\Sigma_p \Omega_{\rm K} \RH^2$\rev{,} is $3.9\ 10^{-3}$ \ME yr$^{-1}$ \rev{and} nearly identical to the actual, 3-D gas mass flux.

\rev{The accretion rates for the largest size bin ($\sim$0.3--1 cm) are approximately $3\, 10^{-6}\,\MEyr$, $7\,10^{-6}\,\MEyr$ and $2\,10^{-5}\,\MEyr$, for the three embryo masses, respectively (cf.  \Fig{accretion_size_distribution}). The accretion rates scale, as expected, approximately as the Hill rate time the Stokes number (proportional to $M^{2/3} a^{-3/4}$ with MMSN scalings).}

\Fig{acc_eff_m095} illustrates the \rev{local} efficiency of accretion of particles across particular spheres in the \texttt{m095t10} simulation. \rev{The `local accretion efficiency' is defined as the net accretion rate through a sphere of radius $R$, normalised by the accretion rate obtained by counting only the solids transport inwards through the same sphere \citep[note that this definition differs from other uses of "accretion efficiency" in the literature; cf.][]{Lambrechts2012,Guillot2014,Ormel2017b}.}
Evidently, a large fraction of all the particles entering the Hill sphere leave again without being accreted. \rev{As illustrated in \Fig{acc_eff_m095}, the local efficiency is particularly low (near 0\%) for the lowest mass particles, which are well coupled to the gas and, due to temporal fluctuations in the accretion rate, can in fact temporarily show negative values.} The local efficiency\rev{, meanwhile,} rapidly grows for smaller radial distances, and at $\sim$30$\rp$ the local efficiency is nearly 100\%, except for particle sizes below 0.003 cm, which are still very well coupled to the gas there. \rev{Over size bins where the efficiencies are statistically well determined, they also scale approximately linearly with size. This is a consequence of the denominator (the unsigned, radially inward transport rate) being approximately independent of particle size (at a given radius) in our regime.}

We do not see significant systematic differences when different pressure bump parameters $\zeta_p$ are used for the same mass embryo. This is to be expected, since the results will only change when the radial settling has become significant. One would expect on the one hand a tendency for  reduction of the accretion rates, because of the smaller $y$-velocities closer to the orbital radius of the embryo, and on the other hand a tendency for increase, due to the accumulation of large size particles.  Ultimately, the accretion rates would be constrained by the radial influx of particles from larger orbital radii into the pressure trap.

\begin{figure*}
  \centering
  \subfigure{\includegraphics[width=0.33\textwidth]{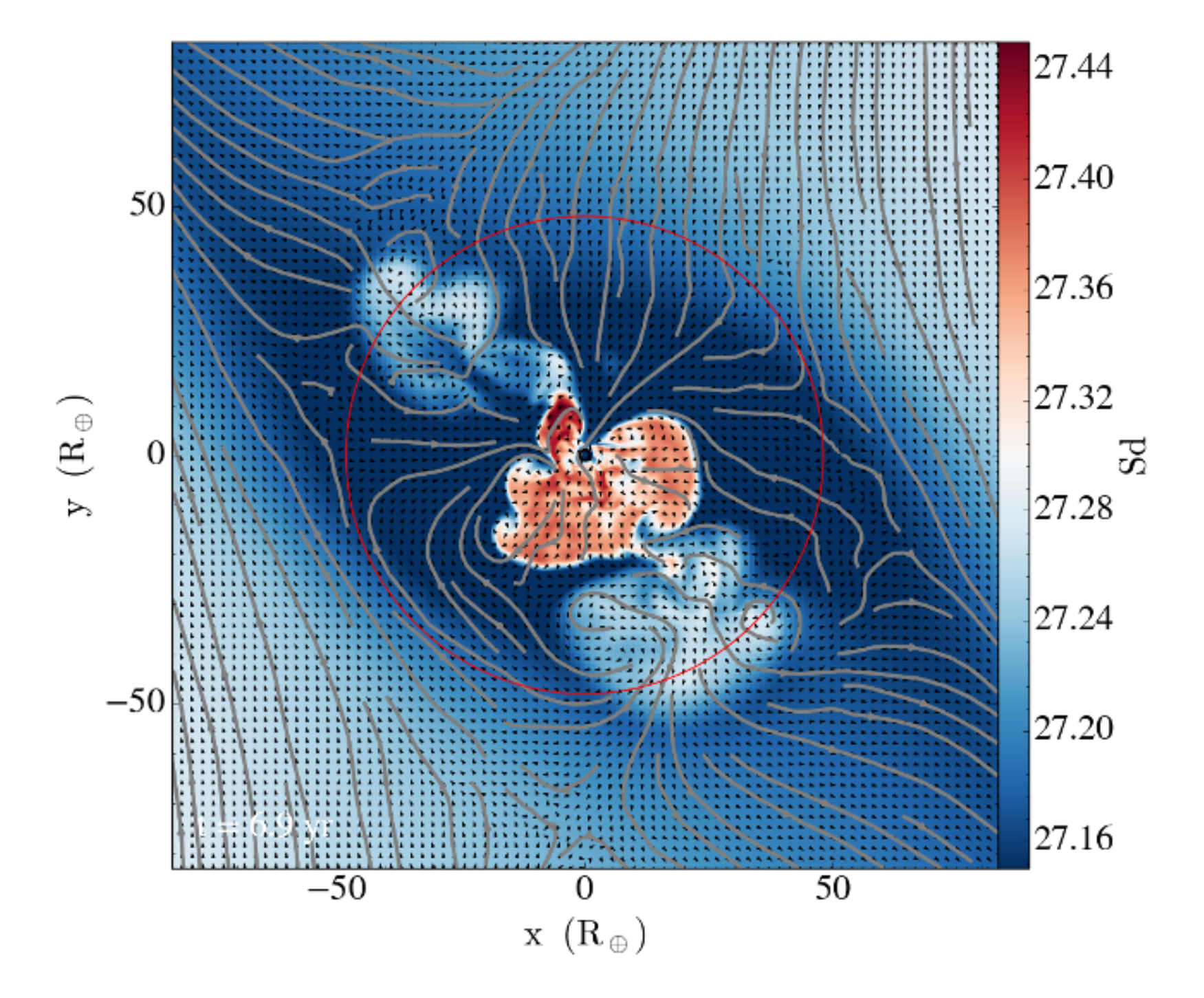}}\hfill
  \subfigure{\includegraphics[width=0.33\textwidth]{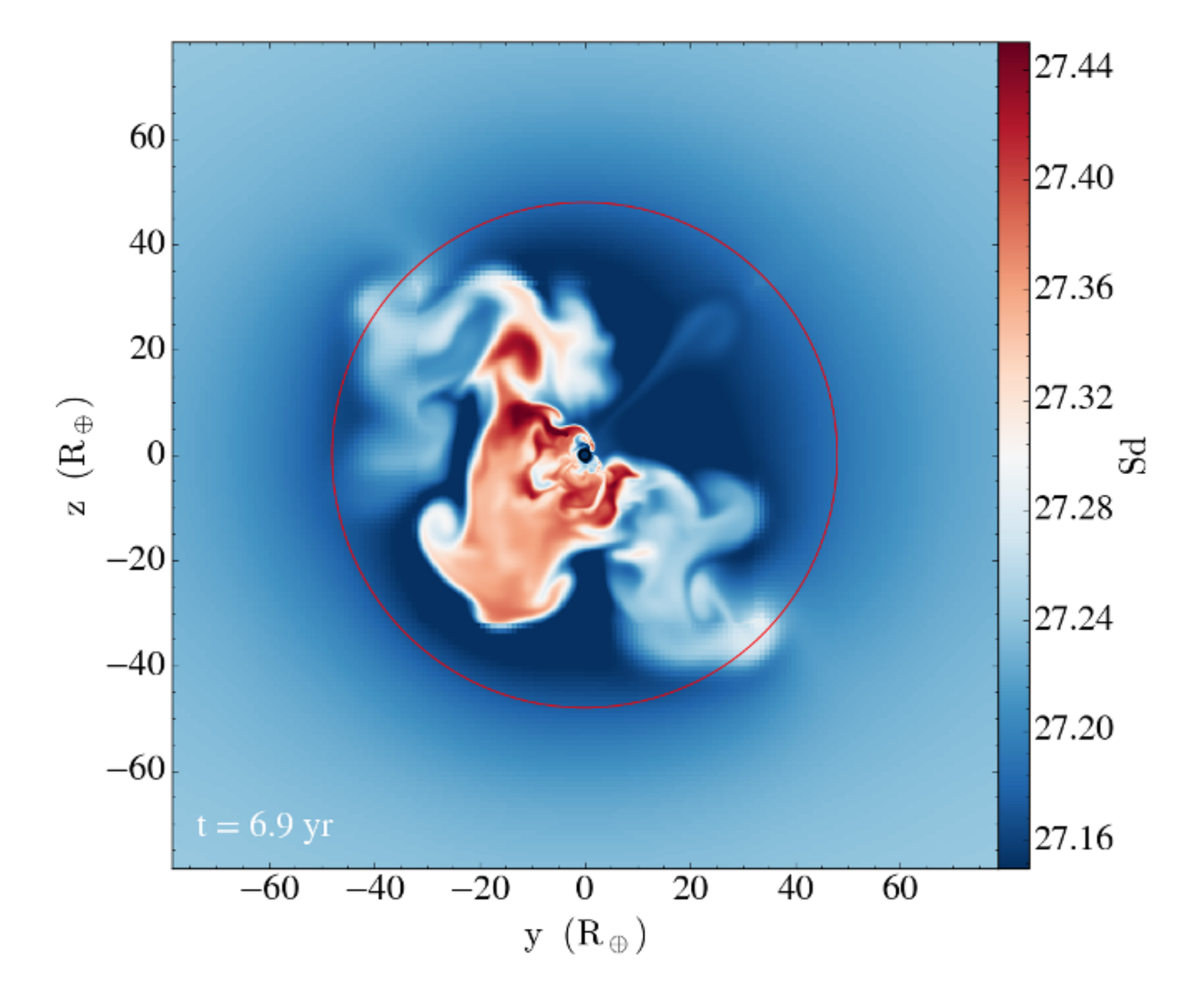}}\hfill
  \subfigure{\includegraphics[width=0.3\textwidth]{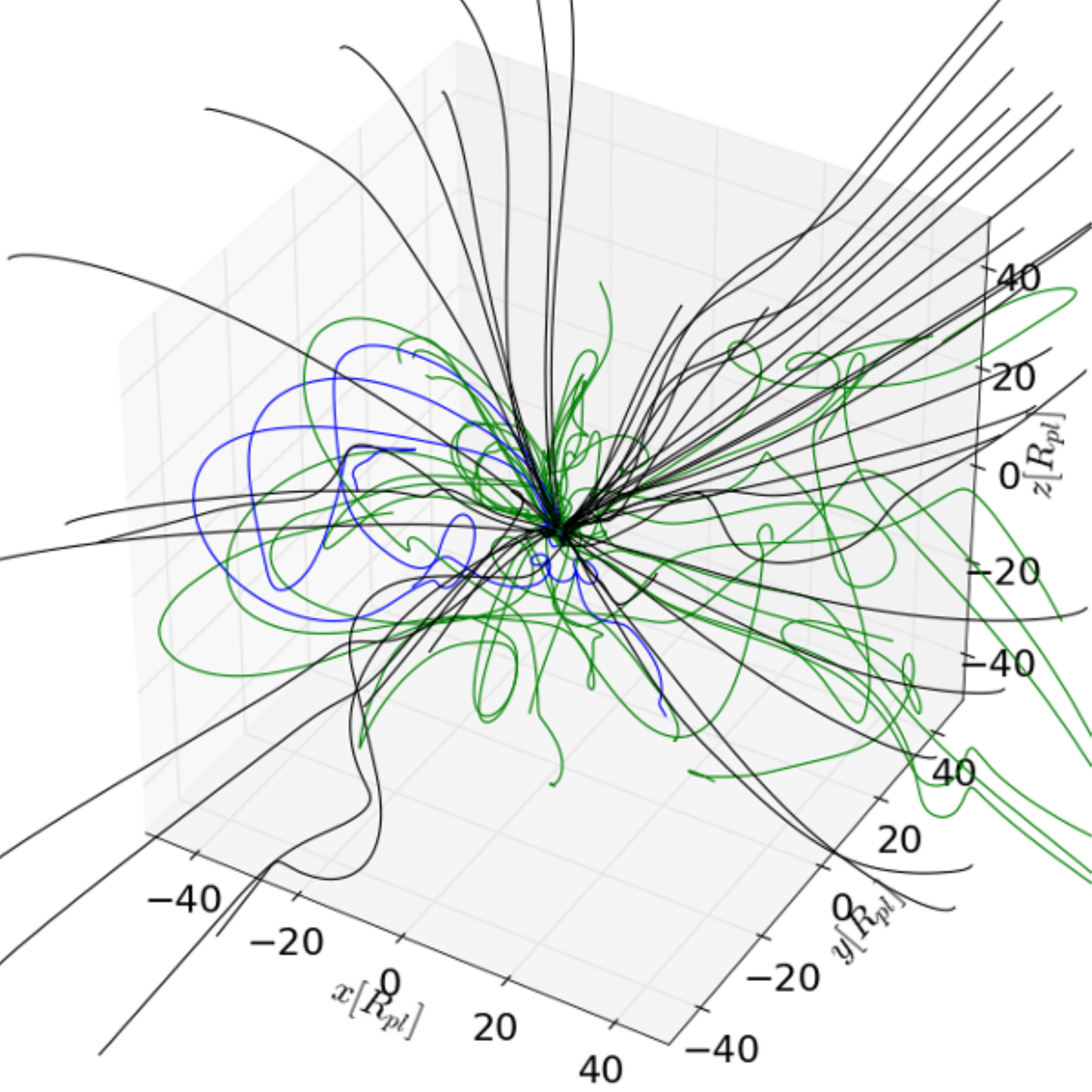}}\hfill
  \caption{From left to right: $xy$ slice in the midplane, $yz$ slice at $x$ = 0 of the entropy per unit mass and particle paths in the vicinity of the core in a single snapshot of accretion driven convection in the \texttt{m095t10-conv} run. For clarity, in the middle panel, the gas streamlines are omitted. In the right panel, black curves show traces for particles of size $0.1 < s \leq 1$ cm, blue $0.01 < s \leq 0.1$ and green $0.001 < s  \leq 0.01$ cm, respectively.}
  \label{fig:accretion_conv}
\end{figure*}

\section{Accretion driven convection}
\label{sec:convection}
\def\Qa{Q_{\rm accr}}
The high resolution close to the embryo surfaces enables us to resolve convection driven by accretion heating. Accretion of solids onto the embryo releases a lot of potential energy, which is converted to heat via the friction force. To simulate the effects of the heating due to the accretion, we added an extra heating term $\Qa$ to one of our simulations, \texttt{m095t10-conv}. Considering a shell of extent $dr$ at some distance $r$ from the embryo, the volume $dV = A dr$, where $A= 4\pi r^2$, then has a potential energy difference (which is per unit mass) given by $\Phi(r+dr) - \Phi(r)$.  To get an energy per unit volume we need to multiply with the mass accreted per unit time:
\begin{equation}
\Qa = \frac{\dot{M}G M_\mathrm{p}}{4\pi r^4},
\label{eq:accr_heat}
\end{equation}
where $\Qa$ is the heating rate per unit volume and $\dot{M}$ is the mass accretion rate (which we set to 10$^{-6}\ \MEyr$). The heating rate, $\Qa$, is then added to the entropy equation (Eq.\ \ref{eq:entropy}) via:
\begin{equation}
\frac{\partial s}{\partial t} = ... + \frac{\rho \Qa}{P_\mathrm{gas}}.
\label{eq:entropywithheating}
\end{equation}
\Fig{accretion_conv} illustrates the consequences of the addition of the accretion heating term in the \texttt{m095t10-conv} run. It shows extensive convective patterns, which considerably modify the gas flows around the embryo (left panel). The horseshoe orbits have somewhat receded, and the weak random flows that appear in non-convective runs are effectively overwhelmed by the much stronger convective flow patterns. The middle panel shows a vertical projection of the convective motions, which extend out to nearly 40 embryo radii. The vicinity of the embryo is no longer in a quasi steady state - the gas flow patterns constantly change, and the interconnection between the primordial atmosphere and the disk is thus strengthened; i.e., no parts of the atmosphere are effectively isolated. The convection is very dynamic, with local velocities reaching $\sim$1 km s$^{-1}$, inducing sound waves when the convective cells reach the outskirts of the atmosphere. The direction of the outflows are constantly changing and now lack any significant constant patterns, but do have an approximately isotropic distribution of the velocity dispersion.

When we see or model convection at a stellar surface (e.g.\ in the form of `granulation' on the Sun), the cellular patterns that develop are very much constrained by conservation of mass. In a stratification where mass density drops very rapidly with height, mass conservation forces the flows to ``turn over'', from up to down, which again forces the horizontal size to be limited (lest the ratio of horizontal to vertical speed become very large). Here, we have a situation that is quite different, with a ``small'' body at the centre of a large spherical volume. This situation allows the flows much greater freedom, with a possibility of expansion in all directions. 

The heating driving the convection in this study, computed assuming $\dot{M}=10^{-6}$ \ME yr$^{-1}$, is not consistent with the actual, about 20 times larger measured accretion rates (e.g.\ \Fig{accretion_size_distribution}). Moreover, radiative energy transfer and a more realistic EOS would tend to change the properties of convection, so to reach true consistency, improvements are necessary. Nevertheless, it is possible to learn already from the current experiment, which can illustrate to what extent the convective motions are able to affect accretion rates relative to an analogous non-convective case.

\begin{figure}
\centering
    \includegraphics[width=\columnwidth]{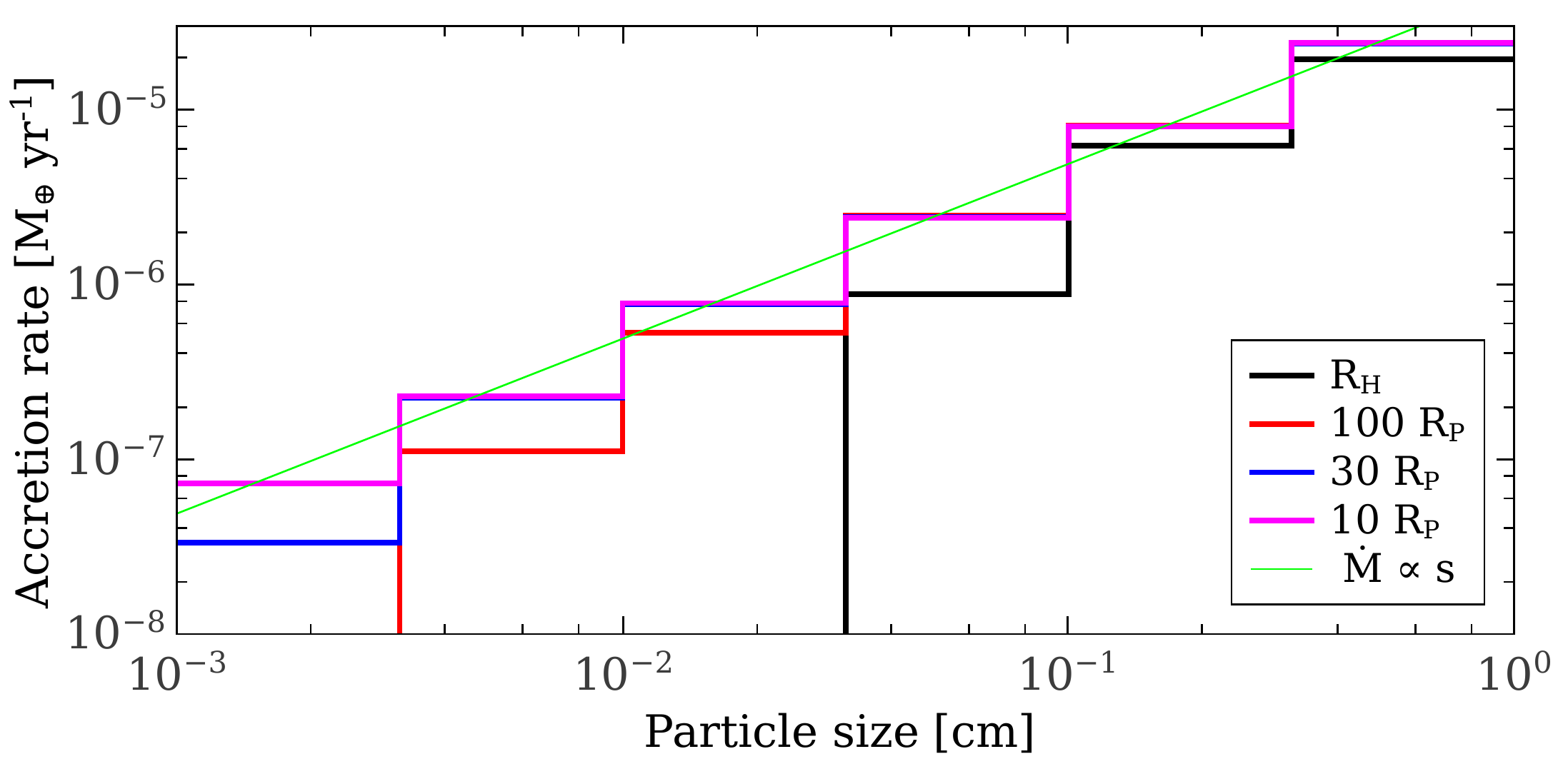}
    \caption{Pebble accretion rates \rev{in the} accretion driven convection \rev{run} (\texttt{m095t10-conv}). Particle sizes are split into 6 representative logarithmic size bins. The green line shows a mass accretion rate proportional to particle size, $s$.}
    \label{fig:m095_conv_rates}
\end{figure}

\begin{figure}
\centering
    \includegraphics[width=\columnwidth]{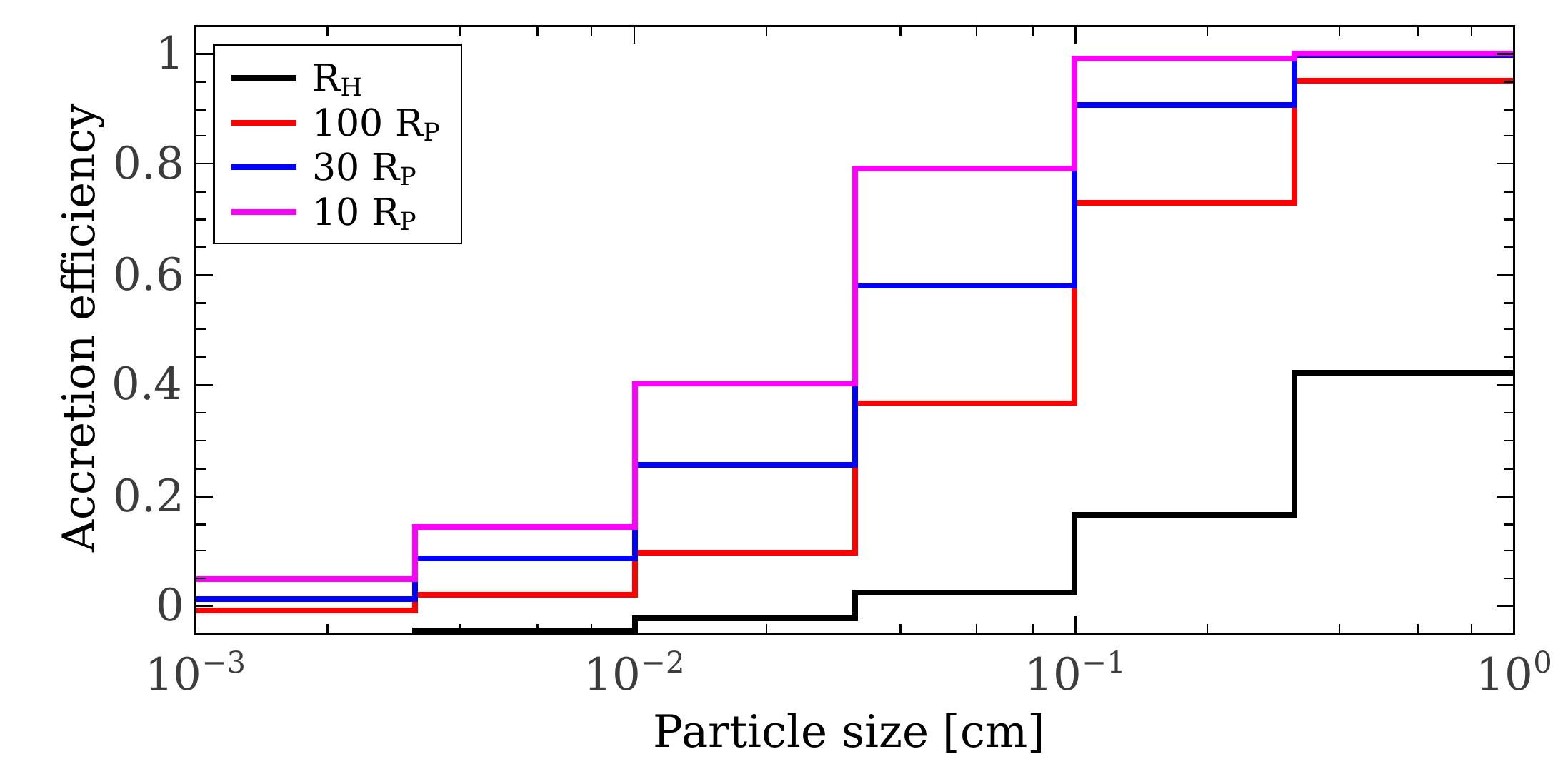}
    \caption{\rev{Local} pebble accretion efficiency \rev{in the} accretion driven convection \rev{run} (\texttt{m095t10-conv}). Particle sizes are split into 6 representative logarithmic size bins.}
    \label{fig:m095_conv_eff}
\end{figure}

The right hand panel of \Fig{accretion_conv} shows a small subset of high cadence paths for accreted particles, colour coded by size. Although their paths are affected, the largest particles are not significantly stirred up by the convective motions. The smaller the particles are, however, the more they are affected by the convective motions. Quantitative measurements show that the average of the unsigned mass fluxes through spherical surfaces where convection is active is much larger than in the non-convective case.

\begin{figure}
\centering
    \includegraphics[width=\columnwidth]{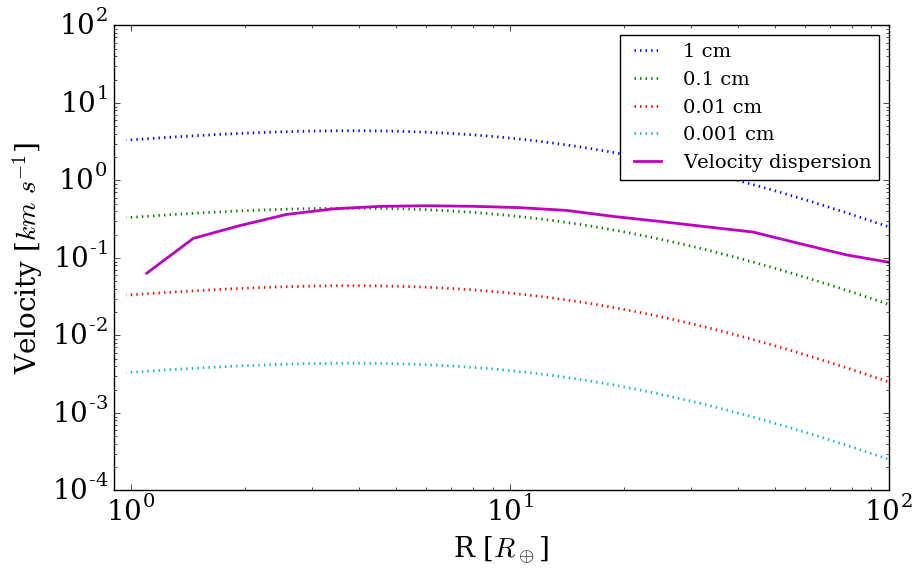}
    \caption{Free fall drag velocities of particles of size 1 cm (blue), 0.1 cm (green), 0.01 cm (red) and 0.001 cm (cyan). The velocity dispersion of the \texttt{m095t10-conv} run is shown in magenta.}
    \label{fig:ffdv}
\end{figure}

However, as illustrated in \Fig{m095_conv_rates}, we find that convection in the current case does not significantly reduce the net accretion rate \rev{onto the embryo}. Formally, the \rev{local} accretion efficiency (the net \rev{mass} accretion rate as a fraction of the unsigned\rev{, radially inward} solid mass flux\rev{; \Fig{m095_conv_eff}}) is reduced significantly for \rev{smaller} particles. \rev{This is because convective motions gives rise to larger unsigned rates of radial mass transport, and hence increase the denominator in the definition of local accretion efficiency. The convective motions also cause larger temporal fluctuations, and as a result we measure negative net accretion rates and a local accretion efficiency of less than zero at $\RB$ for some of the smallest size bins.} The convective velocities in this case are generally on the order of the free fall drift velocities of the particles, as illustrated in \Fig{ffdv}, but the systematic drift of particles relative to the gas continues, and since the convective motions do not result in a net transport of mass, their effect on the particle drift is a second-order effect. It is possible, however, that this could change if the convective motions were sufficiently resolved and included more turbulent, small-scale motions.

The effect on the particle accretion rate is also likely to increase, and possibly become quite significant, when accretion heating is included self-consistently by including radiative energy transfer and a realistic EOS. The resulting, self-consistent accretion rates are likely to be lower than the current values, but the convective motions are likely to be stronger than in our current, nominal case.

\section{Summary and Conclusions}
\label{sec:conclusions_discussion}

In this paper, we have reported the results from high resolution, nested \rev{grid}, 3D hydrodynamic and particle simulations which focus on solid accretion onto Mars- to Earth-mass planetary embryos. The primary goals of this work were to establish realistic three-dimensional flow patterns inside and outside the Hill sphere, to use these to determine accurate accretion rates, and to investigate how the accretion rates depend on embryo mass, pebble size, and other factors.

Our excellent spatial resolution and efficient time integration, using locally determined time steps, made it possible---using only a single, 20-core compute node per run---to perform the first-ever simulations of pebble accretion covering scales from several disk scale heights to a few percent of a planet radius, including atmospheres that self-consistently connect to a surrounding accretion disk. It is also the first time that accretion heating driven convection has been modelled in a domain with such a large dynamic range.

We conducted a series of simulations with different planetary embryo masses and strengths of a pressure trap. To be able to accurately measure accretion rates, and to study how they are affected by convection, we injected millions of macro-particles with representative sizes ranging from 10 micron to 1 m. The motion of the particles were followed, taking into account drag forces relative to the gas, while feedback from the particle ensemble on the gas was neglected---consistent with the modest dust-to-gas ratios occurring in the models. The motions of a subset of particles were recorded with high cadence\rev{, allowing for their trajectories and histories to be studied in great detail.}

Our main results may be summarized as follows:
\begin{enumerate}
  \item Starting from approximately hydrostatic initial conditions, relaxation periods on the order of one orbit are sufficient to establish near quasi-stationary conditions in the vicinity of the Hill sphere.
  \item The relaxed atmospheric structures do not differ dramatically from spherically symmetric hydrostatic atmospheres matched to surrounding disk conditions at the Hill radius. 
  \item Adiabatic and hydrostatic atmospheres with an ideal gas and $\gamma$ = 1.4 do not differ substantially from atmospheres computed with a more realistic, tabular EOS \citep{Tomida2016}.
  \item The presence of convection changes conditions well inside the Hill sphere from near-steady flows to dynamically evolving flows, albeit stationary statistics (e.g.\ velocity dispersion as a function of radius) develop in less than an orbital period.
  \item The convective motions result in an enhanced mass exchange between different layers inside the Hill sphere, but do not significantly affect the mass exchange through the Hill sphere with the surroundings, nor do they affect the net accretion rates significantly.
  \item In our fiducial 1/10 MMSN model, the (unsigned) gas mass flux through the Hill sphere for our 3D horseshoe flows is about $4\ 10^{-3}\ \MEyr$, and thus agrees closely with the canonical ``Hill rate'', $\Sigma \Omega \RH^2 = 3.9\ 10^{-3}\ \MEyr$ (obtained assuming $\Sigma = 170$ g cm$^{-2}$ and $\RH/R_{\oplus}=231$).
  \item The gas mass flux, scaled with the dust-to-gas ratio, is reflected in similar rates of particles passing both in and out of the Hill sphere, with inflow increasingly dominate for increasing particle size.
  \item As illustrated by Figs.\ \ref{fig:pebbles_m01t00d11_midplane} and \ref{fig:accretion_m095t10d10p}, the particles that are not accreted and leave the Hill sphere have preferred paths along the separators of the horseshoe flow and are concentrated towards the midplane. 
  \item In contrast, at the canonical Bondi sphere, the particle fluxes are mainly inward (except for the smallest particle sizes), and hence changes in the detailed structure of the layers close to the embryo (e.g.\ due to the EOS, including from evaporation of pebbles) are not important for the final accretion rates.
  \item The accretion rates for particles \revtwo{is observed to} scale linearly with particle size, \revtwo{different than for unperturbed shear flow conditions, where it scales as $s^{2/3}$ \citep{Ormel2010,Lambrechts2012,Lambrechts2014}. We attribute the steepening of the power-law to aerodynamic deflection of well-coupled particles under the higher densities and very different flow patterns in our simulations relative to unperturbed shear flow conditions.}
  \item We find that the accretion rate of 0.3 -- 1 cm particles for an 0.95 \ME\, embryo immersed in a 1/10 MMSN disk, assuming a dust-to-gas ratio of 0.01 and assuming that the dust has settled to a midplane layer with height $H < R_\mathrm{H}$, is about $2\ 10^{-5}$ \ME yr$^{-1}$.  Since the accretion rates scale essentially linearly with particle size and inversely with gas density, the accretion rate for the same size particles would be essentially the same in higher density disks; the larger supply of solids in a denser disk will tend to be cancelled by a correspondingly slower accretion speed.
  \item For our 1/10 MMSN solar mass disk, the accretion rate saturates (no longer increases with particles size) for particles larger than about 1 m, while for denser disks the particles can be even larger. Thus, only for such large particle sizes will the accretion rates onto an embryo increase with the disk surface density. Conversely, if mm-size pebbles dominate the mass budget, the accretion rates depend only weakly on the disk surface density.
\end{enumerate}

Having such a high spatial resolution means that we could only cover a limited parameter space and run the simulations for a relatively small number of orbital periods. We have also relied on an ideal gas EOS, while in reality the high temperatures near the embryo surface would lead to the dissociation of molecular hydrogen (e.g.\ \citealt{dangelo2013}), and therefore result in a much denser and somewhat cooler inner atmosphere. This would not significantly alter the accretion efficiency of the larger particles, however, as these effects occur at small radii from which these particles are not able to escape.

Given the accretion rates determined in Section 5.3, we can estimate the growth time scales from low mass seeds to full planets, under the specified conditions (i.e.\ a gas surface density of 170 g cm$^{-3}$, a dust-to-gas ratio of 0.01, and solids from within $H < \RH$. As per the results discussed above, to lowest order, the effects of disk density on mass supply and accretion speed cancel. The remaining dominant scaling is the dependence on embryo mass, which is approximately $M^{2/3}$, corresponding to a mass that grows approximately as $t^3$. This implies that, as long as this approximate scaling persists, growth from a small seed takes on the order of three times the instantaneous mass divided by the instantaneous accretion rate.

We conclude that if growth from low mass embryos to full size planets is dominated by accretion of 0.3 -- 1 cm size particles, the accretion from a low mass seed would have taken of the order of 0.15 million years in the case of Earth, and of the order of 0.1 million years in the case of Mars. With chondrule size particles---if we take them to be 0.3-1 mm---the growth times are 10 times longer.  These estimates assume a local ratio of dust-to-gas surface densities of 0.01. Increased dust-to-gas ratios, e.g., due to radial accumulation of solids in a pressure trap, would lead to correspondingly reduced accretion times.

During the growth phase, embryos would inevitably be surrounded by hot, primordial atmospheres of much larger mass than the current day atmospheres of the rocky Solar System planets. The decrease in disk densities required to stop pebble accretion would also result in removal of much of the primordial atmosphere as it expanded to match the decreasing pressure at its outer boundary \citep{Nordlund2011, Schlichting2015,Ginzburg2016}. The isotopic signatures present in noble gases in the Earth's atmosphere may provide indirect evidence related to this \citep{Pepin1991,Pepin1992a,Pepin1992b}.

A decrease in gas and dust densities will lead to increased rates of radiative cooling, at first dominated by cooling in the vertical direction. The consequential, gradual flattening of the hydrostatic structure inside the Hill sphere could lead to the formation of a circumplanetary disk with prograde rotation \citep[cf.][]{johansenlacerda2010}. In order to model such an evolution, it is crucial to realistically include the major sources of heating, energy transport, and cooling: Accretion heating, the resulting convection, and radiative cooling while also using a realistic EOS (e.g.\ \citealt{Tomida2013}; cf.\ Fig.\ \ref{fig:eos}) and opacities.

To this end, we are currently conducting simulations that include these effects. In the future, we will also model the accretion heating in a manner consistent with the solid accretion rate (cf.\ Sect.\ \ref{sec:convection}) and include pebble destruction via ablation (e.g.\ \citealt{brouwers_how_2017,alibert_maximum_2017}).

\section*{Acknowledgements}

We  would  like  to  thank  the anonymous referee for in-depth and helpful comments that have helped to significantly improve the quality of this manuscript. The work of AP and \AA N was supported by grant 1323-00199B from the Danish Council for Independent Research (DFF). CWO is supported by the Netherlands Organization for Scientific Research (NWO; VIDI  project 639.042.422). The Centre for Star and Planet Formation is funded by the Danish National Research Foundation (DNRF97). Storage and computing resources at the University of Copenhagen HPC centre, funded in part by Villum Fonden (VKR023406), were used to carry out the simulations presented here. The authors are grateful to Kengo Tomida and Yasunori Hori, who kindly provided the tabular EOS. AP is grateful to the Anton Pannekoek Institute at the University of Amsterdam for a three month stay during which part of this work was carried out. This research made use of the YT project for analysis and visualisation \citep{Turk+2011}.




\newpage
\bibliographystyle{mnras}
\bibliography{ms} 



\appendix
\section{Additional figures}
\label{app:additional_figures}

\begin{figure*}
  \centering
  \subfigure[Midplane.]{\includegraphics[width=0.33\linewidth]{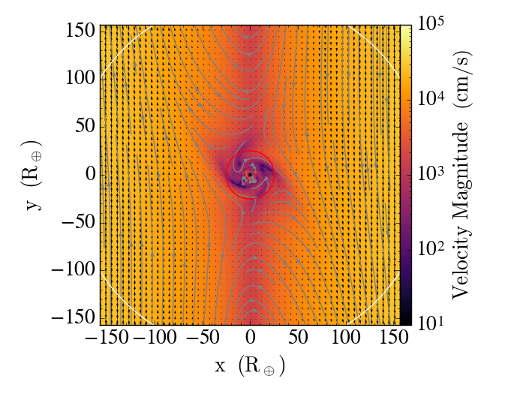}}\hfill
  \subfigure[$z$ = 10 $\rp$.]{\includegraphics[width=0.33\linewidth]{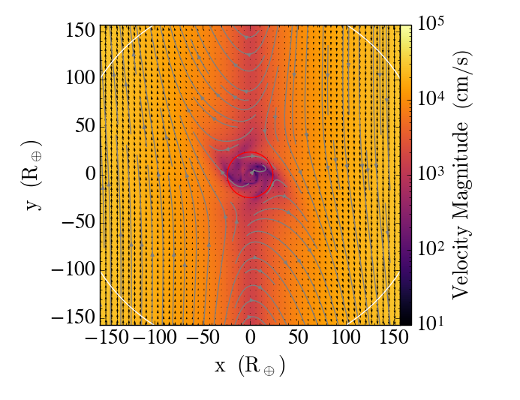}}\hfill
  \subfigure[$z$ = 25 $\rp$.]{\includegraphics[width=0.33\linewidth]{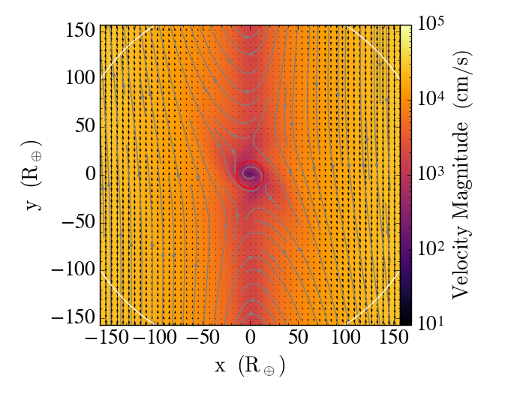}}\\
  \subfigure[$z$ = 75 $\rp$.]{\includegraphics[width=0.33\linewidth]{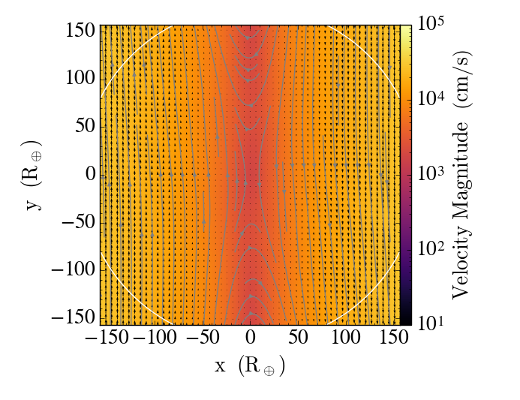}}\hfill
    \subfigure[Vertical component of the vorticity in the midplane.]{\includegraphics[width=0.33\linewidth]{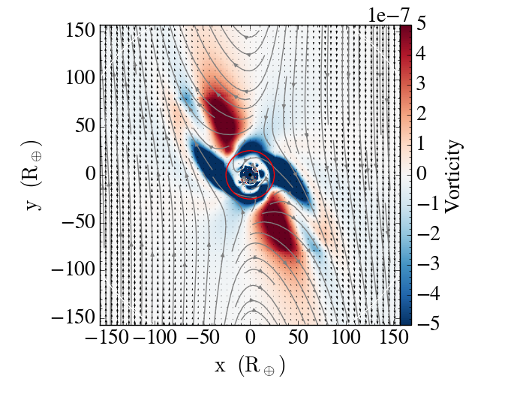}}
      \subfigure[$xz$ slice of the vertical velocity component at $y$ = 0.]{\includegraphics[width=0.33\linewidth]{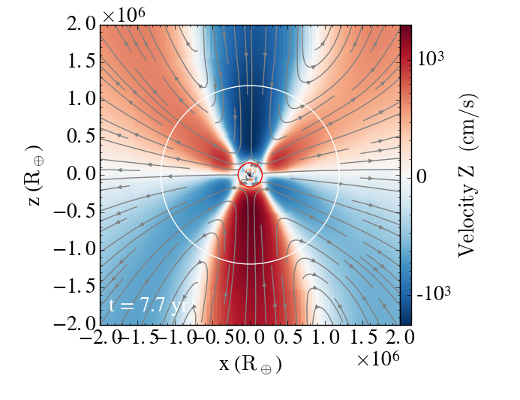}}
  \caption{Same as in \Fig{gas_high_mass}, but for the $M$=0.5 \ME\, embryo from the \texttt{m05t00} run.}
  \label{fig:gas_medium_mass}
\end{figure*}

\Fig{gas_medium_mass} shows the details of the gas flow in the \texttt{m05t00} run for the $M$ = 0.5 \ME\ planetary embryo. The similarity with the higher mass case is apparently because embryos of such masses have well established primordial atmospheres, which are very close to being hydrostatic, leading to very similar flows in the vicinity of the Hill sphere.

\begin{figure*}
  \centering
  \includegraphics[width=0.33\textwidth]{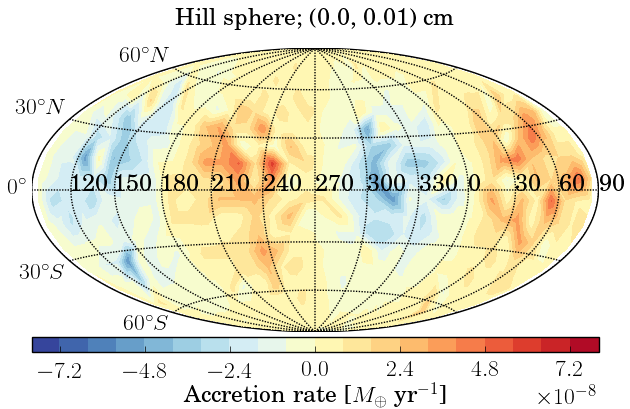}\includegraphics[width=0.33\textwidth]{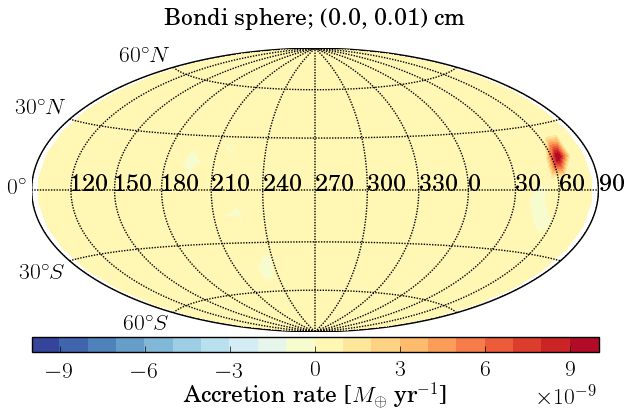}\includegraphics[width=0.33\textwidth]{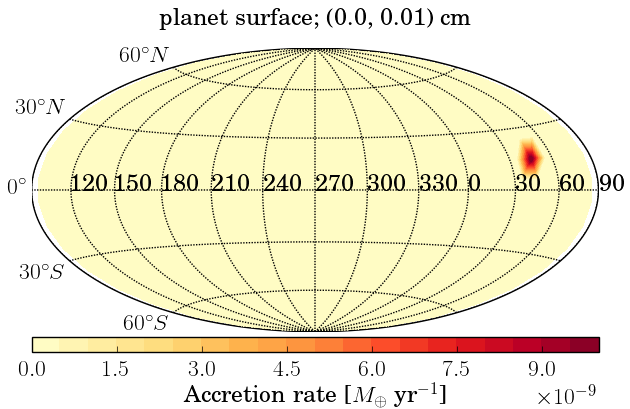}
  \includegraphics[width=0.33\textwidth]{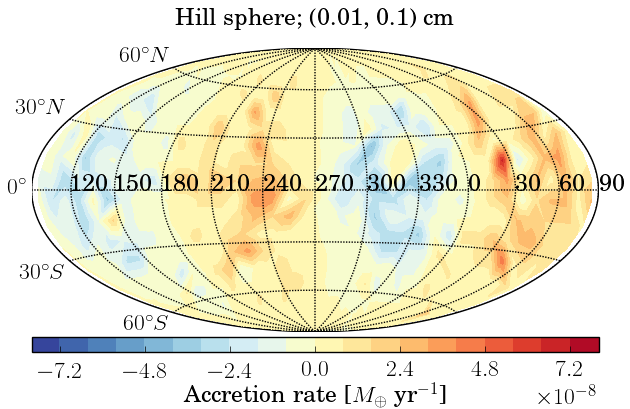}\includegraphics[width=0.33\textwidth]{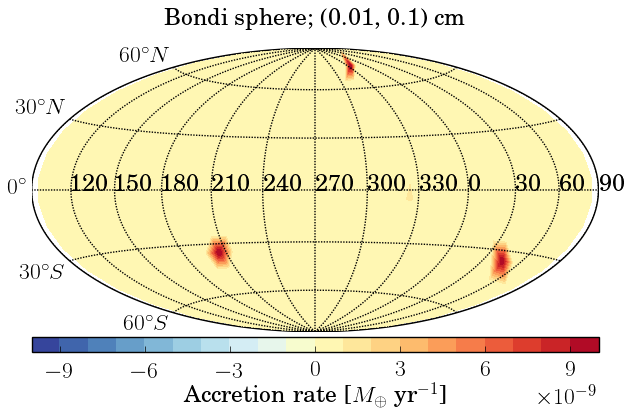}\includegraphics[width=0.33\textwidth]{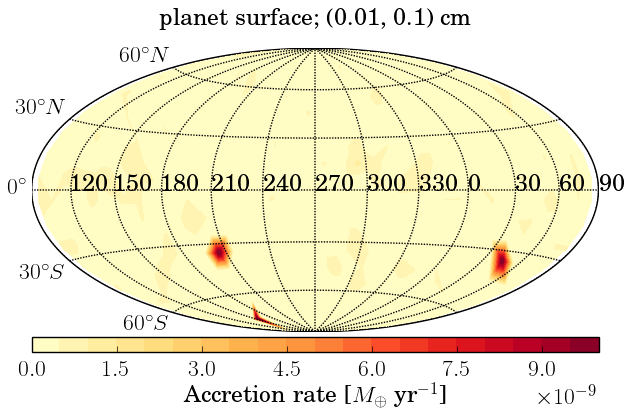}
  \includegraphics[width=0.33\textwidth]{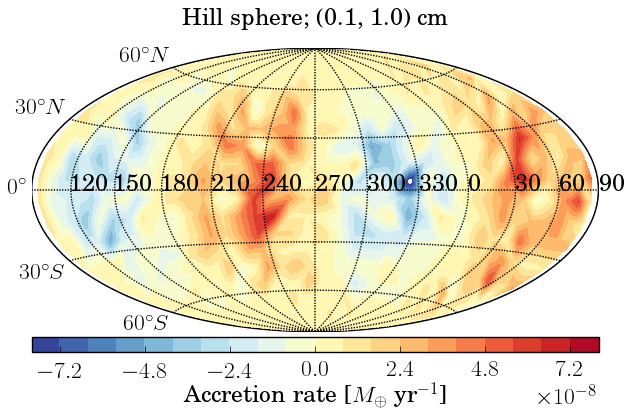}\includegraphics[width=0.33\textwidth]{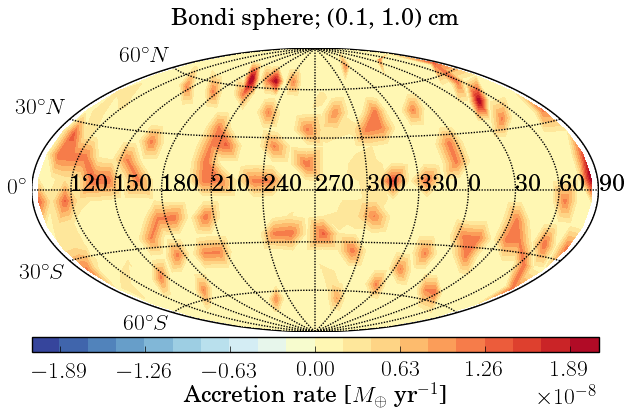}\includegraphics[width=0.33\textwidth]{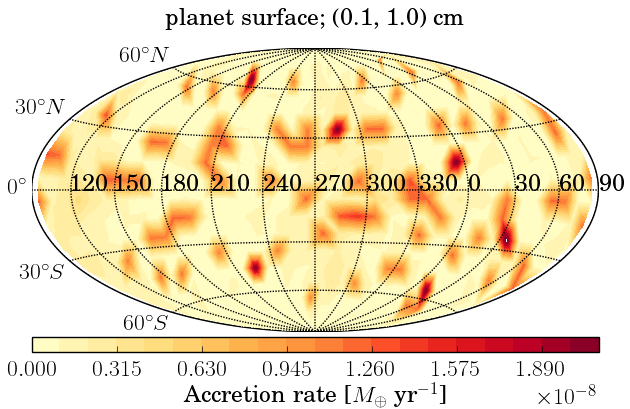}
  \caption{Same as \Fig{accretion_m095t10d10p}, but for the \texttt{m05t10} run.}
  \label{fig:accretion_m05t10p}
\end{figure*}

\Fig{accretion_m05t10p} shows the mass accretion rates over the course of the \texttt{m05t10} run. The accretion rates are about a factor of two lower than in the \texttt{m095t10} case, and the preferred accretion paths through the Hill sphere are once again closer to the midplane and the stagnation points (separators). 

\begin{figure*}
  \centering
  \includegraphics[width=0.33\textwidth]{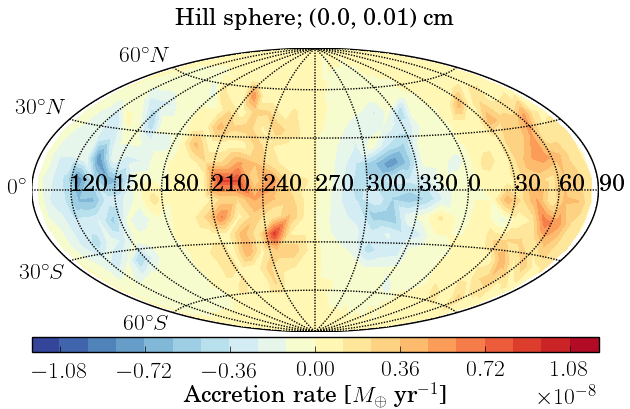}\includegraphics[width=0.33\textwidth]{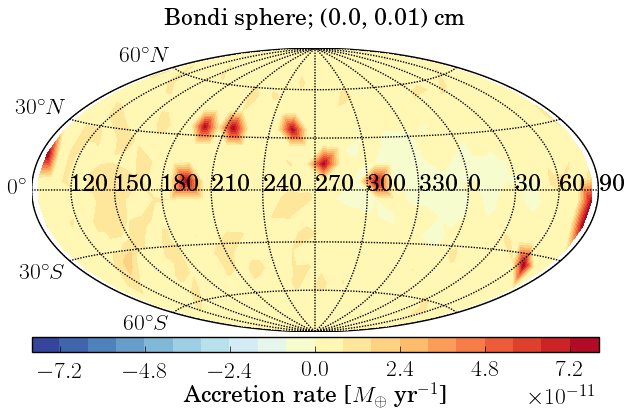}\includegraphics[width=0.33\textwidth]{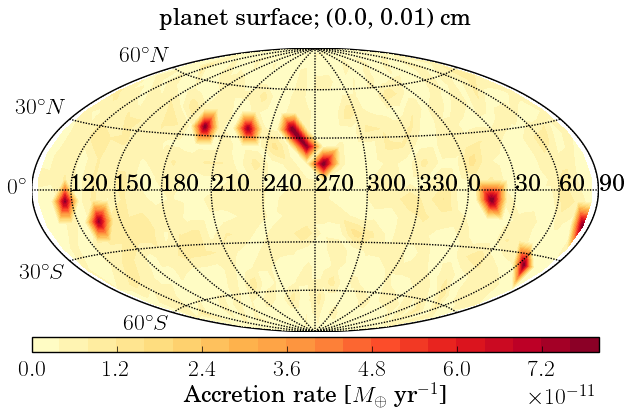}
  \includegraphics[width=0.33\textwidth]{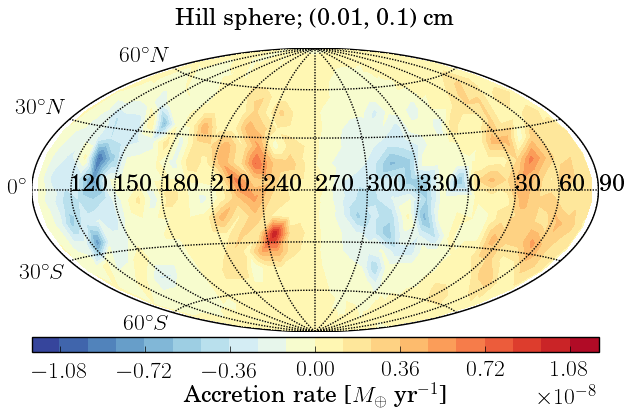}\includegraphics[width=0.33\textwidth]{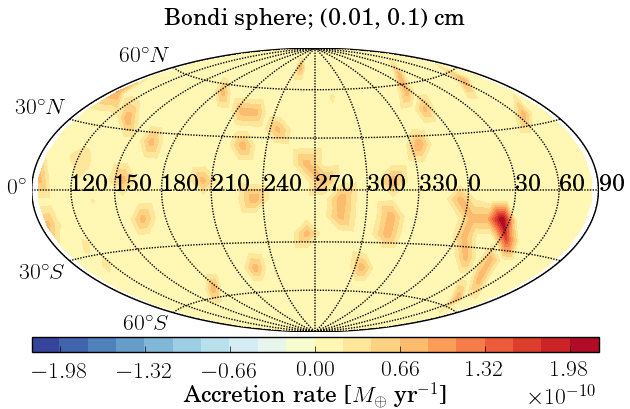}\includegraphics[width=0.33\textwidth]{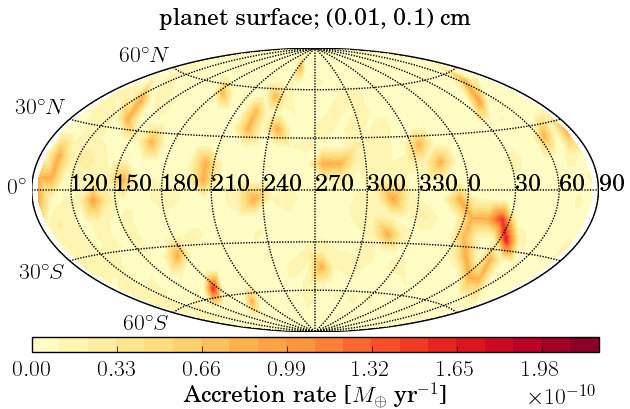}
  \includegraphics[width=0.33\textwidth]{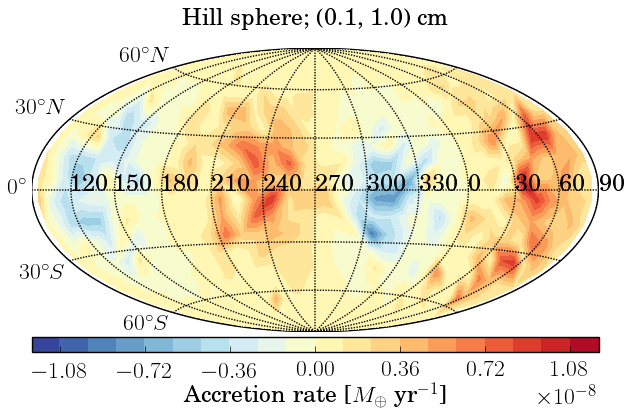}\includegraphics[width=0.33\textwidth]{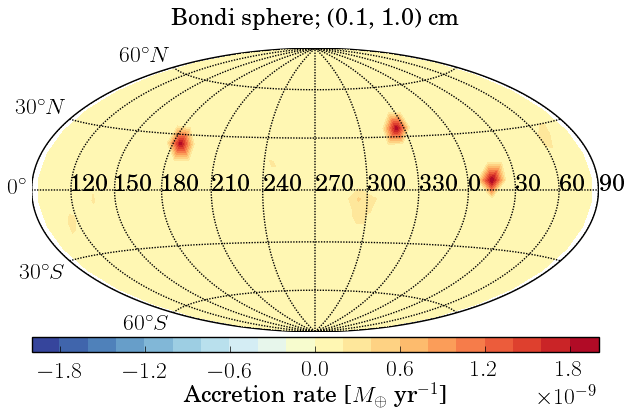}\includegraphics[width=0.33\textwidth]{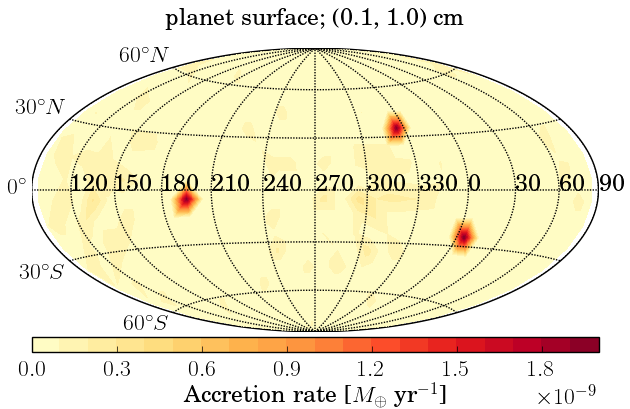}
  \caption{Same as \Fig{accretion_m095t10d10p}, but for the \texttt{m01t10} run.}
  \label{fig:accretion_m01t10p}
\end{figure*}

\Fig{accretion_m01t10p} shows the mass accretion rates over the course of the \texttt{m01t10} run. Note that the contour levels are modified, in order to enhance the features. The simulation show\rev{s} a similar scenario to the cases with higher mass embryos, except the mass accretion rates are once again smaller. Also, there is \rev{effectively} no \rev{loss of} mass \rev{from} $\RB$. This is because the $\RB$ for such a small mass embryo is just $\sim$12.3 embryo radii, \rev{or just 4\% of $\RH$, and therefore actually deeper in the potential well of the embryo than in the higher mass cases}.

\bsp	
\label{lastpage}
\end{document}